\documentclass[
 aps,
 pra,
 twocolumn,
 superscriptaddress,
 letterpaper,
 amsmath,
 amssymb
]{revtex4-1}

\usepackage{graphicx}
\usepackage{dcolumn}
\usepackage{bm}
\usepackage[colorinlistoftodos]{todonotes}
\usepackage{tikz}

\usepackage{lipsum}

\usepackage{physics}

\usepackage{xcolor}
\usepackage[normalem]{ulem}
\usepackage{xr-hyper}
\usepackage[breaklinks]{hyperref}

\hypersetup{
    colorlinks=true,
    linkcolor=red,
    filecolor=magenta,      
    urlcolor=cyan,
}

\begin{document}

\title{Dynamically Reconfigurable Photon Exchange in a Superconducting Quantum Processor}
\author{Brian Marinelli}
\thanks{These two authors contributed equally. Correspondence should be addressed to brian\_marinelli@berkeley.edu and jie.roger.luo@gmail.com}
\affiliation{Quantum Nanoelectronics Laboratory, University of California, Berkeley, Berkeley CA 94720}
\affiliation{Computational Research Division, Lawrence Berkeley National Laboratory, Berkeley CA 94720}
\author{Jie Luo}
\thanks{These two authors contributed equally. Correspondence should be addressed to brian\_marinelli@berkeley.edu and jie.roger.luo@gmail.com}
\affiliation{Quantum Nanoelectronics Laboratory, University of California, Berkeley, Berkeley CA 94720}
\affiliation{Computational Research Division, Lawrence Berkeley National Laboratory, Berkeley CA 94720}
\affiliation{Anyon Computing Inc., Emeryville CA 94608}
\author{Hengjiang Ren}
\affiliation{Anyon Computing Inc., Emeryville CA 94608}
\author{Bethany M. Niedzielski}
\affiliation{Lincoln Laboratory, Massachusetts Institute of Technology, Lexington, MA 02421-6426}
\author{David K. Kim}
\affiliation{Lincoln Laboratory, Massachusetts Institute of Technology, Lexington, MA 02421-6426}
\author{Rabindra Das}
\affiliation{Lincoln Laboratory, Massachusetts Institute of Technology, Lexington, MA 02421-6426}
\author{Mollie Schwartz}
\affiliation{Lincoln Laboratory, Massachusetts Institute of Technology, Lexington, MA 02421-6426}
\author{David I. Santiago}
\affiliation{Quantum Nanoelectronics Laboratory, University of California, Berkeley, Berkeley CA 94720}
\affiliation{Computational Research Division, Lawrence Berkeley National Laboratory, Berkeley CA 94720}
\author{Irfan Siddiqi}
\affiliation{Quantum Nanoelectronics Laboratory, University of California, Berkeley, Berkeley CA 94720}
\affiliation{Computational Research Division, Lawrence Berkeley National Laboratory, Berkeley CA 94720}
\affiliation{Materials Science Division, Lawrence Berkeley National Laboratory, Berkeley CA 94720}

\date{\today}

\begin{abstract}
    Realizing the advantages of quantum computation requires access to the full Hilbert space of states of many quantum bits (qubits). Thus, large-scale quantum computation faces the challenge of efficiently generating entanglement between many qubits. In systems with a limited number of direct connections between qubits, entanglement between non-nearest neighbor qubits is generated by a series of nearest neighbor gates, which exponentially suppresses the resulting fidelity. Here we propose and demonstrate a novel, on-chip photon exchange network. This photonic network is embedded in a superconducting quantum processor (QPU) to implement an arbitrarily reconfigurable qubit connectivity graph. We show long-range qubit-qubit interactions between qubits with a maximum spatial separation of $9.2~\text{cm}$ along a meandered bus resonator and achieve photon exchange rates up to $g_{\text{qq}} = 2\pi \cross 0.9~\text{MHz}$. These experimental demonstrations provide a foundation to realize highly connected, reconfigurable quantum photonic networks and opens a new path towards modular quantum computing.
\end{abstract}

\maketitle

Rapid development of quantum information processing platforms in recent years has shown the promise of utilizing programmable QPUs to carry out nontrivial digital and analog quantum programs that could be scaled into practical applications in the near future \cite{2019GoogleSupremacy,2020GoogleEnergies,2021GoogleQAOA,fedorov2022,OMalley2016,Islam2013,seetharam2021,johri2021,Nam2020,Gill2020,Kivlichan2018}. However, almost all near-term nontrivial quantum programs require entanglement operations that can be carried out on a fully connected network of high coherence qubits \cite{Koch2020,fedorov2022,svore2006,Korenblit2012,Vaidya2018,Wright2019,Manovitz2020,Periwal2021,Bluvstein2022,Hamerly2019,Altman2021,Nandkishore2015,Joshi2020,Islam2013,Landig2016,Ebadi2021,Joshi2022}. 
As quantum devices are expected to steadily scale up, qubit interconnectivity is one of the major challenges to achieve quantum advantage in the noisy intermediate-scale quantum (NISQ) era \cite{fedorov2022}.
Gate decomposition by concatenation of many imperfect two-qubit gates exponentially suppresses the resulting gate fidelity \cite{Koch2020}. This limits the performance of near-term quantum programs according to the connectivity available in the hardware. Superconducting qubit based hardware platforms have shown promise for quantum applications but current state of the art devices are limited to 2D connectivity graphs \cite{2019GoogleSupremacy, Jurcevic_2021,fedorov2022}. Meanwhile, due to their high degree of connectivity, ion-trap QPUs have successfully performed quantum simulations that would require additional resources in superconducting qubit based devices \cite{Zhang2017, Landsman2019}. 

Combining the high cooperativity and controllability of superconducting QPUs with reconfigurable (up to all-to-all) connectivity into a single platform would allow for the execution of new classes of quantum algorithms, simulations, and other tasks on NISQ era QPUs. Recent demonstrations representing progress in this direction include a Molmer-Sorenson gate on up to four transmons \cite{Lu_MSGate} and a metamaterial waveguide based quantum simulator \cite{zhang_metamaterial}. Additionally, there have been recent proposals for all-to-all connectivity using a tunable bus based on flux qubits \cite{NoriFluxQubitBus} or a fixed bus with coupling controlled by single qubit AC flux drives \cite{Onodera2020}. The critical component in each proposal is a central coupling element that allows for controllable coupling and higher connectivity between distant qubits. Here we propose and demonstrate an architecture that utilizes a common, flux tunable, multi-mode resonator (Bus) to parametrically exchange photons between multiple superconducting qubits and thus creating an on-chip parametric photon exchange network between qubits for the first time. This allows for an arbitrary and dynamically reconfigurable qubit connectivity graph which is programmed at the level of the room temperature frequency multiplexed microwave controls. 


In the following we present a description of the circuit element we propose to realize the photon-exchange mediating Bus. The model is then used to fit the measured Bus spectrum to extract the circuit parameters. The analysis is extended to the case of time periodic modulation of the Bus external flux bias where a Fourier space analysis predicts qubit-Bus parametric coupling rates that agree well with the measured coupling rates. We then show that desired spectrally selective qubit-qubit photon exchange can be generated. We also experimentally characterize the time dynamics generated by these qubit-qubit interactions, show programmable directed photon routing between three qubits, and realize long-range photon exchange between pairs of qubits separated by up to 9.2~cm.

\section*{Tunable CPW Bus Resonator \label{sec:tunableresonator}}
\begin{figure*}
    \centering
    \includegraphics{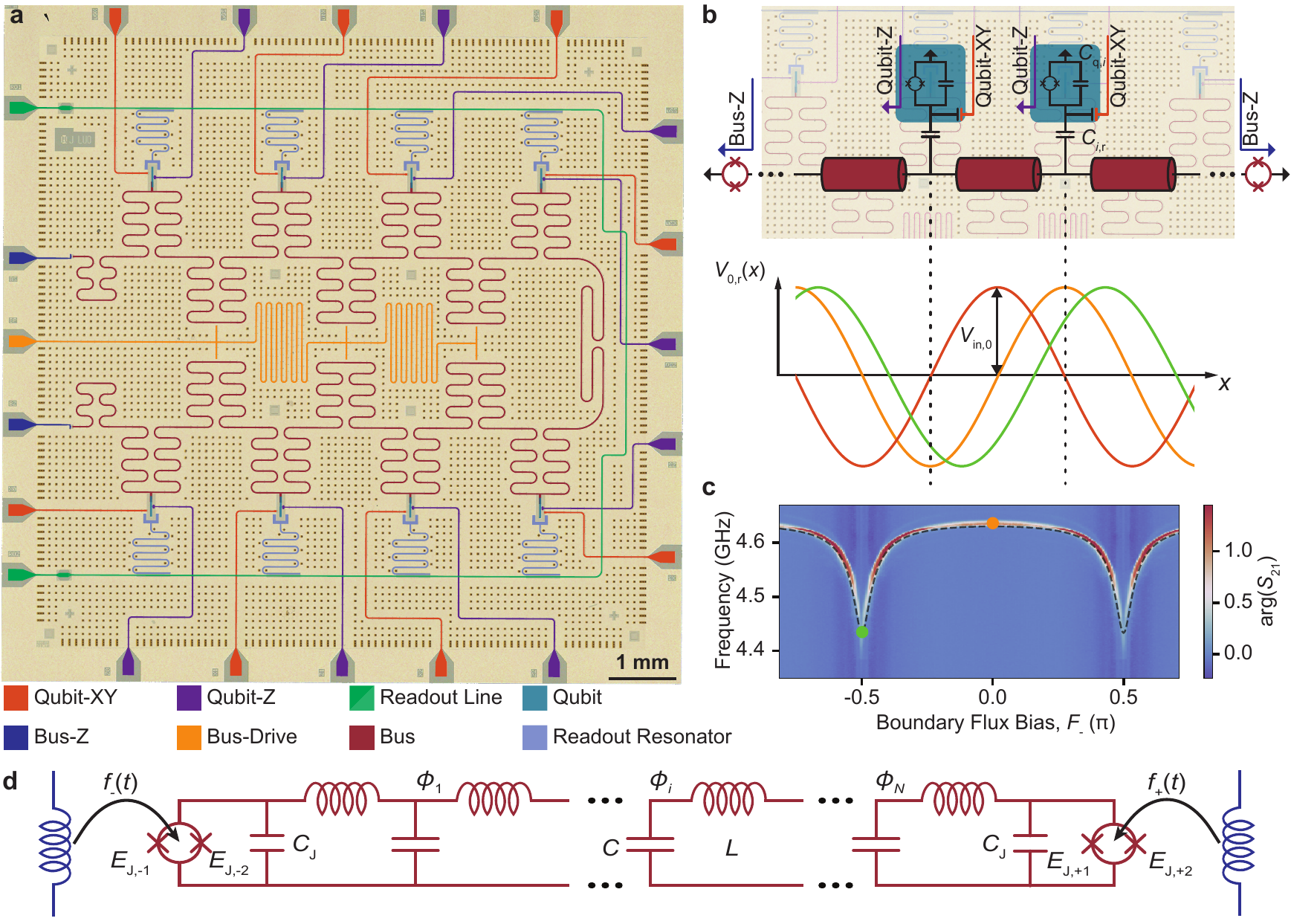}
    \caption{\textbf{a}, False colored, composite microscope image of the fabricated device. The device was fabricated through a flip-chip process with the qubit chip containing qubits and readout circuitry and the wiring chip containing control lines and the Bus \cite{Rosenberg20173D, Yost2020}. Qubits are labelled starting with Q$_{1}$ in the top left and wrapping around along the trace of the Bus (red) to Q$_{8}$ on the bottom left. \textbf{b}, Schematic showing two qubits coupled to the Bus resonator in the upper figure. In the lower figure, the Bus coupling mode wave profile seen by qubits for effective open-open boundaries (red), effective short-short boundaries (yellow), and effective open-short boundaries (green) as a function of position $x$ along the Bus. \textbf{c}, The measured Bus spectrum for the 7th mode as a function of flux bias at one boundary. The dots correspond to the bias points for profiles of the same colors in \textbf{b}. See SI Note~\ref{sec:freqdomain} for measurement details. The multi-mode spectrum was used to extract Bus parameters through fitting to a double-side SQUID terminated CPW resonator model. \textbf{d}, Effective lumped LC circuit model for the Bus resonator constructed with a long coplanar waveguide terminated by SQUIDs at both ends. The magnetic flux through each SQUID, $f_{\pm}(t)$ is controlled by the Bus-Z flux line next to it.} 
    \label{fig:spidernetcircuit}
\end{figure*}

Linear coplanar waveguide (CPW) resonators are commonly used in superconducting quantum circuits to mediate coupling between transmon qubits \cite{ChowCR, Paik_RIPGate, Lu_MSGate}. The qubits are coupled capacitively to the CPW resonator in which case the qubit-resonator coupling, $\hbar g_{0i}=C_{i,\text{r}}V_{0,\mathrm{r}}(x_{i})V_{0,i}$ for qubit $i$, is determined by the zero point fluctuation voltage amplitude of the resonator mode profile $V_{0,\mathrm{r}}(x_{i})$ at the qubit position $x_{i}$, the qubit zero point voltage fluctuation $V_{0,i}=\sqrt{\frac{\hbar\omega_{\text{q},i}}{2C_{\text{q},i}}}$ for qubit frequency $\omega_{\text{q},i}$ and shunt capacitance $C_{\text{q},i}$, and capacitance between the qubit and resonator, $C_{i,\text{r}}$. The effective resonator mediated qubit-qubit coupling is approximately $g_{ij}=g_{i}g_{j}(1/\Delta_{i} + 1/\Delta_{j})$ where $\Delta_{i} =\omega_{\text{q},i}-\omega_{\text{r}}$ is the detuning between the frequencies of qubit $i$ and the resonator. Since $g_{ij} \sim V_{0,\mathrm{r}}(x_{i})V_{0,\mathrm{r}}(x_{j})$, if the field profile along the resonator can be tuned then the qubit-qubit coupling can also be tuned. 

For a CPW resonator, the mode profile will be determined by its geometry and boundary conditions. We use a resonator with tunable boundary conditions to mediate the qubit-qubit coupling \cite{CasparisVoltageTunableBus}. In this work, tunability is introduced by terminating the CPW resonator to ground through superconducting quantum interference devices (SQUIDs) \cite{Sandberg2008, Palacios-Laloy2008, Castellanos-Beltran2007} at both boundaries (see Fig.~\ref{fig:spidernetcircuit}\textbf{b}). The SQUIDs act like flux-tunable boundary impedances, $Z(\Phi)=\sqrt{L_{\mathrm{SQ}}(\Phi)/C_{\mathrm{SQ}}}$ where $L_{\mathrm{SQ}}(\Phi)=L_{\text{SQ}}(0)/\cos \Phi$ is the flux tunable Josephson inductance of the symmetric SQUID and $C_{\mathrm{SQ}}$ is the SQUID capacitance to ground. Let $Z_{0}$ be the impedance of the bulk CPW resonator, then proper design of the boundary SQUID allows for tuning the boundary condition from short, $Z(\Phi) \ll Z_{0}$, to open, $Z(\Phi) \gg Z_{0}$ \cite{JohanssonCasimir1,JohanssonCasimir2}. If the qubits are positioned at nodes (antinodes) of the mode profile with both boundary conditions tuned to open, then they will be positioned at antinodes (nodes) of the mode profile with both boundary conditions tuned to short, as depicted schematically in Fig.~\ref{fig:spidernetcircuit}\textbf{b}. As a result the qubit-resonator and qubit-qubit couplings can be tuned from nearly zero to a maximum value determined by flux-tunable circuit parameters. 

In an attempt to formalize the above intuition we now present an explicit circuit model of the tunable resonator. The circuit diagram is shown in Fig.~\ref{fig:spidernetcircuit}\textbf{d}. For concreteness, the length of the resonator is $\ell$ with the position along the resonator $-\ell/2 \leq x \leq \ell/2$. The capacitance and inductance per unit length in the bulk of the CPW are $C_{0}$ and $L_{0}$ respectively, giving a bulk CPW impedance of $Z_{0}=\sqrt{L_{0}/C_{0}}$. The junctions composing the boundary SQUIDs have capacitances $C_{si}$ and Josephson energies $E_{\mathrm{J},si}$ where $s=\pm$ denotes the $x= \pm \ell/2$ boundary and $i=1,2$ labels the junctions within each SQUID. The dynamical variables are the bulk node fluxes $\phi_{j}$ for $j=0,\dots,N$ and the boundary SQUID node fluxes $\phi_{s}$ for $s=\pm$. The time dependent external fluxes threading the boundary SQUIDs are $f_{s}(t)$. Our model for the system and subsequent calculations follow closely previous work in \cite{WustmannParametricResonance} and \cite{WallquistSelectiveCoupling} but are generalized to the case where both ends of the tunable resonator are terminated by SQUIDs to yield the equations of motion below. The details of this calculation can be found in SI Note~\ref{sec:busEOM}.

The bulk phase field, $\phi(x,t)$, satisfies
\begin{equation}
    \frac{\partial^{2}\phi}{\partial t^{2}}-v^{2}\frac{\partial^{2}\phi}{\partial x^{2}}=0
    \label{eqn:eombulk}
\end{equation}
which is simply the standard electromagnetic wave equation where $v=1/\sqrt{L_{0}C_{0}}$ is the speed of light in the bulk CPW. At the boundaries the phase field satisfies 
\begin{equation}
    \lim_{x \rightarrow \pm \frac{\ell}{2}}\left[\frac{2}{\omega_{\mathrm{J}}^{2}}\frac{\partial^{2}\phi}{\partial t^{2}}+2\cos f_{\pm}(t)\sin\phi \pm \eta d\frac{\partial \phi}{\partial x}\right]=0
    \label{eqn:eombndry}
\end{equation}
where $\omega_{\mathrm{J}}=\sqrt{2E_{\mathrm{C}}E_{\mathrm{J}}}/\hbar$ is the plasma frequency of the Josephson junctions composing the boundary SQUIDs, $E_{\mathrm{C}}=(2e)^{2}/2C_{\mathrm{J}}$ is the junction charging energy, and $\eta=E_{\mathrm{L}}/E_{\mathrm{J}}$ is the ratio of the bulk inductive energy $E_{\mathrm{L}}=\left(\frac{\Phi_{0}}{2\pi}\right)^{2}/\ell L_{0}$ to the boundary Josephson energy, $E_{\mathrm{J}}$. For simplicity we assume the SQUIDs are symmetric and that the SQUIDs at either boundary are identical. Thus the junction capacitance $C_{\mathrm{J}}=C_{si}$ and $E_{\mathrm{J}}=E_{\mathrm{J},si}$ for all $s$ and $i$. First we consider the case of DC flux bias $f_{\pm}(t)=F_{\pm}$. The resonant frequencies of the flux biased tunable resonator can be solved for numerically and fit to the experimentally measured tunable resonator mode spectrum in order to extract the circuit parameters, as outlined below.

We design and fabricate an eight-qubit version of the proposed architecture, shown in Fig.~\ref{fig:spidernetcircuit}\textbf{a}. The tunable resonator is a $10.55~\text{cm}$, chip-scale meander CPW resonator terminated by SQUIDs at either end. On-chip flux lines (Bus-Z) are used for DC and RF biasing of the boundary SQUIDs and a charge drive line (Bus-Drive) is used to spectroscopically probe the tunable resonator. Eight tunable transmon qubits are capacitively coupled to the Bus resonator. Each qubit has its own on-chip flux (Qubit-Z) and charge (Qubit-XY) control lines and a readout resonator. The eight readout resonators are all coupled to a single feed-line for multiplexed readout. Due to the presence of the chip scale Bus and multiplexed readout line the routing of the numerous control lines becomes a challenge in a planar device. To alleviate these signal routing constraints the device is fabricated using a flip-chip integration process \cite{Rosenberg20173D}. The false colored device image in Fig.~\ref{fig:spidernetcircuit}\textbf{a} is a composite of the microscope images of the two chips overlaid on each other. The qubits and readout resonators are located on the top ``qubit chip'' while the control lines (qubit-XY, qubit-Z, Bus-Z, Bus-drive) and Bus are located on the bottom ``wiring chip''. The readout line originates on the wiring chip (darker green) and is transferred to the qubit chip (lighter green) galvanically by superconducting indium bumps.

The measurement setup allows for observation of the first nine modes of the Bus which are spaced by the free spectral range (FSR) of $\omega_{\text{FSR}}\sim 2\pi \times 660~\text{MHz}$ up to approximately $2\pi \times 6~\text{GHz}$ (see SI Note~\ref{sec:freqdomain}). The bus frequency is tuned by the boundary flux biases, $F_{\pm}$, which tune the boundary conditions. We fix $F_{+}=0$ and sweep $F_{-}$ over the range $-\pi/2 \leq F_{-} \leq \pi/2$ and measure the frequencies $\omega_{n,\text{r}}$ of Bus modes $n=1,\dots,9$ at each bias. Mode 7 and its fit to the theoretical model introduced above are shown in Fig.~\ref{fig:spidernetcircuit}\textbf{c} (extended data over the full frequency range is found in SI Note~\ref{sec:extdataFD}). Modes 3, 5, 6, 7, and 9 are simultaneously fit to a single set of circuit parameters. When the mode frequencies are normalized by the mode index, the fitted modes collapse onto each other while modes 1 and 8 are offset by $\mp 10~\text{MHz}$ so they are excluded from the fit. Modes 2 and 4 can be observed as well but they couple weakly to the qubit used as a spectrometer for these measurements so they are also excluded from the fit. Beyond observing strong agreement between the fit and experimental spectra, we also find that the fit circuit parameters are all within a reasonable range of their designed or simulated values, as given in Table~\ref{tab:DCfitpars}. To obtain a good fit, particularly near the $F_{-}=\pm \pi/2$ bias points, we need to allow for a small SQUID asymmetry, $d=(E_{\mathrm{J},1}-E_{\mathrm{J},2})/(E_{\mathrm{J},1}+E_{\mathrm{J},2})$ in the model. The fit value of $d=0.06$ is within a reasonable range for the typical fabrication variation in the junction critical currents. 

\section*{Parametric Qubit-Bus Coupling \label{sec:modulatedbus}}
\begin{figure*}
    \centering
    \includegraphics{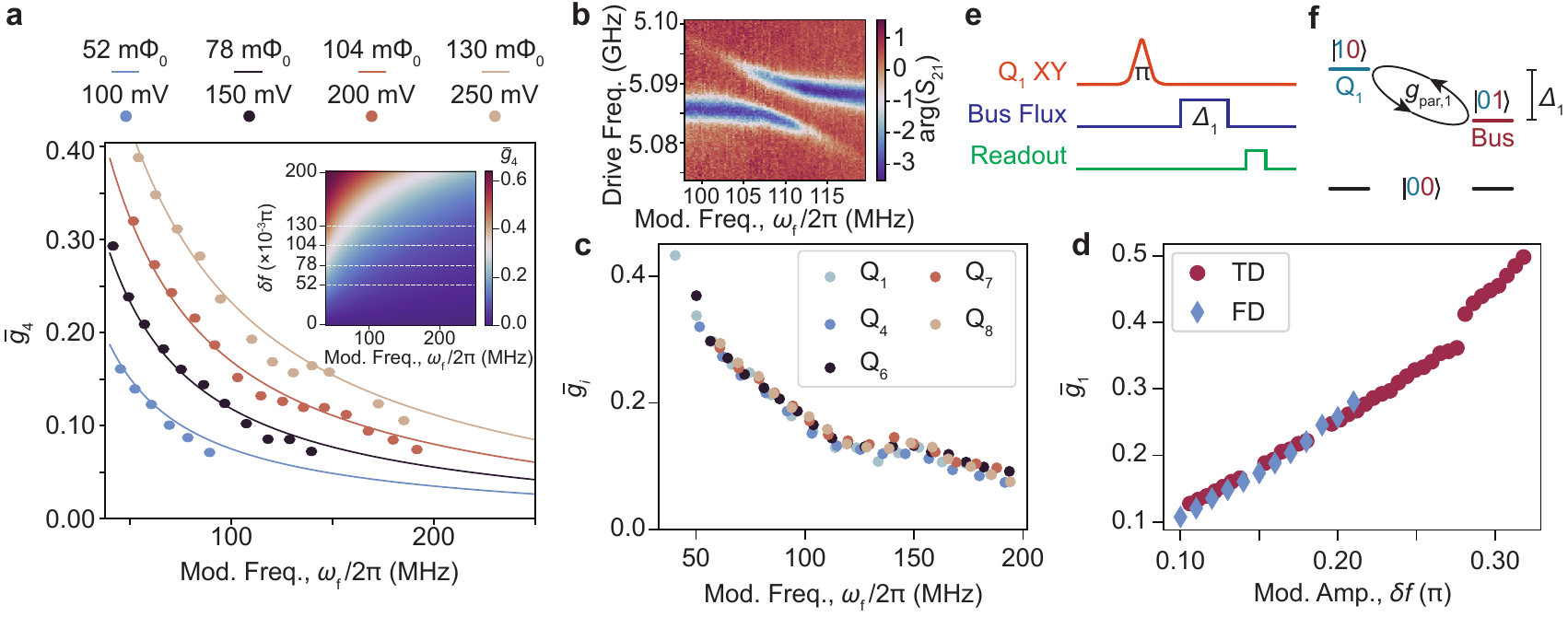}
    \caption{\textbf{a}, Measured (dots) and theory predicted (solid lines) parametric Q$_{4}$-Bus coupling as a function of Bus flux modulation frequency at various modulation amplitudes. The inset of \textbf{a} shows the full 2D theory prediction for $\bar{g}_{4}$ over the relevant parameter ranges with system parameters extracted from fitting the experimental data (see discussion in SI Note~\ref{sec:busmatrixtheory} for details). The dashed white lines in the inset indicate constant modulation amplitude linecuts corresponding to the solid lines in \textbf{a}. \textbf{b}, A typical parametric avoided crossing between Q$_{4}$ and Bus from which the parametric coupling data in \textbf{a} was extracted. \textbf{c}, Parametric couplings between different qubits and the Bus at fixed amplitude ($\delta f=0.104\pi$) and various detunings are extracted from fitting avoided crossings like \textbf{b} and demonstrate that all qubits are coupled symmetrically to the shared Bus as expected. \textbf{d}, Q$_{1}$-Bus parametric coupling rate (red dots) obtained from the photon exchange rate in time domain (TD) and overlaid on the previously obtained parametric coupling rate (blue diamond) from fitting frequency domain (FD) avoided crossings. \textbf{e}, The pulse sequence for measuring time domain coupling rates in \textbf{c}. \textbf{f}, The parametric photon transition process between Q$_{1}$ and Bus at rate $g_{\text{par},1}$ activated by the flux drive to Bus boundaries.}
    \label{fig:parametricqubitbus}
\end{figure*}
In order to generate parametric photon exchange between qubits, we need to apply a time dependent external flux to the Bus boundary SQUIDs. This modulation of the Bus boundary condition modulates the Bus mode profile and produces a time-periodic coupling between the qubits and Bus. To show this more explicitly, we now investigate the response of the resonance mode field profile under periodic boundary flux modulation, $f_{\pm}(t)=F_{\pm} + \delta f_{\pm} \cos(\omega_{\mathrm{f}}t+\psi_{\pm})$. For simplicity, we assume the static bias points of the two SQUIDs are the same, $F=F_{+}=F_{-}$ as well as their modulation amplitudes, $\delta f=\delta f_{+}=\delta f_{-}$. We Fourier transform and linearize the equations of motion, Eqn.~\ref{eqn:eombndry}, and then decompose the Fourier amplitudes into left and right travelling wave components, $\phi_{\pm}(\omega)$. After taking advantage of the system symmetries (details in SI Note~\ref{sec:busmatrixtheory}), we arrive at an infinite dimensional matrix equation for the vector of sideband phase field amplitudes, $\phi_{\pm}(\omega_{p})$,
\begin{equation}
    \label{eq:matrixformBus}
    \left(\mathbf{\Sigma}\pm\mathbf{\Sigma}^{*}\right)\tilde{\mathbf{\Phi}}_{\pm}=0
\end{equation}
where $\mathbf{\Sigma} = \mathbf{\Lambda}(\{e^{-i\frac{1}{2} \pi n}\})\mathbf{M_{+,+}}\mathbf{\Lambda}(\{e^{i\frac{1}{2} \pi n}\})$ is defined in terms of the matrix 
\begin{multline}\label{eq:freqmatrixelementmain}
     [\mathbf{M}_{s=\pm,z=\pm}]_{mp}=\big[(-\alpha\omega_p^2+sz\eta \ell k_p i)\delta_{mp}+\\
     (e^{iF}+(-1)^{m-p}e^{-iF})J_{p-m}(\delta f)e^{z i\frac{\psi_0}{2}(p-m)}\big]e^{sz\frac{\omega_p \ell}{2v}i}
\end{multline}
and $\mathbf{\Lambda}(\{x_{n}\})$ is the diagonal matrix with $n$th diagonal element $x_{n}$. Additionally, $\delta_{mp}$ is the Kronecker-$\delta$ symbol, $\alpha=2/\omega_{\mathrm{J}}^{2}$, $\psi_{0}=\psi_{+}-\psi_{-}$ is the relative phase between the periodic modulation at either boundary, $J_{p-m}$ is the $(p-m)$th Bessel function of the first kind, and $\omega_{p}=vk_{p}=\omega_{\mathrm{r}} + p\omega_{\mathrm{f}}$ is the $p$th sideband frequency with respect to the tunable Bus resonance frequency $\omega_{\mathrm{r}}$. The symmetric($+$)/anti-symmetric($-$) frequency component vector $\tilde{\mathbf{\Phi}}_{\pm}$ is defined such that $\tilde{\mathbf{\Phi}}_{\pm} = \mathbf{\Lambda}(\{e^{-i\frac{1}{2} \pi n}\})(\mathbf{\Phi}_{+} \pm \mathbf{\Phi}_{-})/\sqrt{2}$ where $\mathbf{\Phi}_{\pm} = (\dots,\tilde{\phi}_{\pm}(\omega_{-1}),\tilde{\phi}_{\pm}(\omega_{0}),\tilde{\phi}_{\pm}(\omega_{1}),\dots)$ is a vector of right($+$)/left($-$) propagating wave frequency components. Parametric boundary flux modulation leads to mixing between different sidebands spaced by $\omega_{\text{f}}$ such that the new normal modes of the system are linear combinations of various sidebands.


We solve numerically for the device parameters in Table~\ref{tab:DCfitpars}, obtained from fitting the Bus spectrum to the circuit model (except symmetric SQUIDs, $d=0$, are assumed) after truncating Eqn.~\ref{eq:matrixformBus} to the $D=3$ lowest sidebands. We compute the coupling strength between the Bus and the qubits with the modulation frequency, $\omega_{\mathrm{f}} = \Delta_{i}$ for qubit $i$ (Q$_i$), generating a resonant parametric interaction between the qubit and Bus. After properly normalizing the sideband amplitudes obtained from solving Eqn.~\ref{eq:matrixformBus}, it is straightforward to show that (see SI Note~\ref{sec:busmatrixtheory} for details) the normalized parametric qubit-Bus coupling of Q$_i$ is
\begin{equation}
    \label{eqn:gpar}
    \bar{g}_{i}=\frac{g_{\text{par},i}}{g_{0i}}=\frac{\frac{1}{2}(-1)^{i+1}\left(\tilde{\Phi}^{(1)}_{-}-\tilde{\Phi}^{(-1)}_{-}\right)}{\tilde{\phi}_{\text{in},0}(\omega_{\text{in},\text{r}})}\cos\left(\frac{\omega_{\text{f}}x_{i}}{v}\right)
\end{equation}
where the qubit position along the Bus is $x_{i}$ and the zero point phase field fluctuation of the initial unmodulated Bus mode is $\tilde{\phi}_{\text{in},0}(\omega_{\text{in},\text{r}})$. Thus, the parametric coupling rate is determined by the weights of the Bus sideband amplitudes in the presence of flux modulation.

In order to measure $g_{\text{par},i}$, we perform a three-tone frequency domain (FD) experiment with the Bus biased to $F=\pi/4$ (see SI Note~\ref{sec:freqdomain}). As the phase correlated Bus flux modulation frequency $\omega_{\text{f}}$ is swept through the detuning $\Delta_{i}$ a parametric qubit-Bus avoided crossing is observed between the qubit and the first order Bus sidebands (see Fig.~\ref{fig:parametricqubitbus}\textbf{b}). The size of the avoided crossing is $2g_{\text{par},i}$. We repeat the measurement at multiple modulation amplitudes $\delta f$ and qubit-Bus detunings, $\Delta_{i}$ with the results shown in Fig.~\ref{fig:parametricqubitbus}\textbf{a} for a representative qubit, Q$_{4}$. Data for qubits Q$_{1}$, Q$_{6}$, Q$_{7}$, and Q$_{8}$ can be found in SI Note~\ref{sec:extdataFD}. The circuit parameters are constrained from the fit in Table~\ref{tab:DCfitpars} and $g_{0i}$ is similarly extracted from the avoided crossing observed by tuning Q$_i$ across the 8th order Bus mode. This leaves just a single free parameter in the model, the proportionality constant between the room temperature flux modulation amplitude in volts (V) and the corresponding modulation amplitude in units of $\pi$ after normalizing to the magnetic flux quantum $\Phi_{0}$. We find good agreement between the experimentally measured parametric coupling and the theoretical prediction for a microwave line attenuation of $31~\text{dB}$ which is in good agreement with the cryogenic line and room temperature cable attenuation estimated to be $30.3~\text{dB}$. We attribute the remaining difference to insertion loss in the bias tee used to combine the AC and DC flux biases.

The interaction occurs between the qubits and sidebands of the 8th order Bus mode which is designed to couple symmetrically to all qubits and has the frequency closest to the qubit frequencies. We verify that the qubits all interact symmetrically with the Bus by comparing their parametric couplings for the same modulation amplitude. In Fig.~\ref{fig:parametricqubitbus}\textbf{c} we see that the data all collapse onto a single curve. Based on Eqn.~\ref{eqn:gpar} this qubit position independent coupling is expected in the current case where $\omega_{\text{f}} \ll \omega_{\text{FSR}}$ and $\cos(\omega_{\text{f}}x_{i}/v) \approx 1$.

We corroborate these extensive frequency domain (FD) measurements with time domain (TD) measurements of the qubit-Bus coupling rate. These are performed by first exciting Q$_i$, then modulating the Bus boundary SQUIDs at $\omega_{\text{f}}=\Delta_{i}$ and observing coherent population exchange between the qubit and Bus (see Fig.~\ref{fig:parametricqubitbus}\textbf{e} for the pulse sequence). Repeating this for $\omega_{\text{f}}$ around $\Delta_{i}$ yields the characteristic TD chevron pattern from which the TD coupling rate is extracted from the population exchange rate. Fig.~\ref{fig:parametricqubitbus}\textbf{d} shows that the FD and TD measurements of the qubit-Bus coupling agree well after accounting for an extra $0.5~\text{dB}$ of attenuation from a DC block that was added to the Bus flux lines between the FD and TD measurements. The close agreement validates the frequency domain method for extracting the parametric coupling rates.

\section*{Parametric Qubit-Qubit Coupling \label{sec:qqcoupling}}
Following the demonstration above, we show that the desired direct photon exchange between a pair of qubits can be generated through a time periodic coupling, $g_{i}(t)=g_{0i}+2g_{\text{par},i}\cos(\omega t)$ for $i=1,2$, produced with Bus flux modulation. The static qubit-Bus coupling at a given Bus flux bias $F$ is $g_{0i}$. We arrive at the effective Hamiltonian 
\begin{equation}
    \label{eq:Hinteff}
    H_{\text{int},\text{eff}}/\hbar \approx g_{12}\sigma_{1,-}\sigma_{2,+}+\text{h.c.}
\end{equation}
assuming $\omega=\tilde{\Delta}_{12}$, where $\tilde{\Delta}_{12}=\tilde{\omega}_{\text{q},1}-\tilde{\omega}_{\text{q},2}$ and $\tilde{\omega}_{\text{q},i}$ is the parametrically renormalized frequency of Q$_i$. The details of the calculation can be found in SI Note~\ref{sec:tswt} where we show the effective qubit-qubit coupling rate is
\begin{equation}
    \label{eq:geff}
    g_{12}=\frac{g_{01}g_{02}}{2}\left[\bar{g}_{2}\left(\frac{1}{\Delta_{1}}+\frac{1}{\Sigma_{1}}\right) +\bar{g}_{1}\left(\frac{1}{\Delta_{2}}+\frac{1}{\Sigma_{2}}\right)\right]
\end{equation}
and we define the qubit-Bus sum frequencies $\Sigma_{i}=\omega_{\text{q},i}+\omega_{\text{r}}$. The time periodic coupling to the shared Bus generates an approximate exchange interaction between the qubits.

\begin{figure*}
    \centering
    \includegraphics{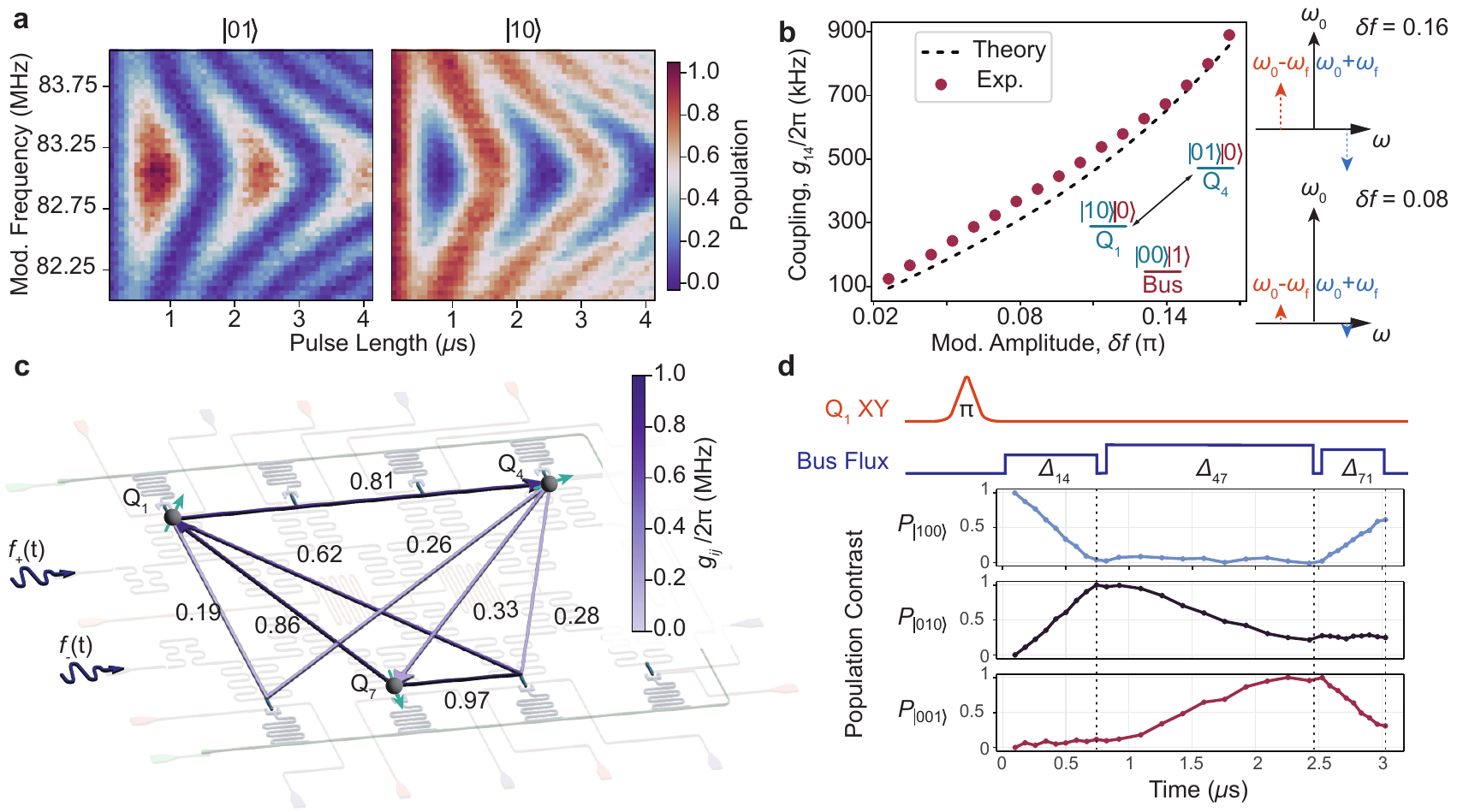}
    \caption{\textbf{a}, Measured coherent photon exchange between Q$_{1}$ and Q$_{4}$ using the pulse sequence in Fig.~\ref{fig:parametricqubitbus}\textbf{e} with the Bus flux pulse having modulation frequency matching the detuning, $\Delta_{14}$, between the qubits. \textbf{b}, The comparison between the experimental qubit-qubit exchange rate (red dots) as a function of modulation amplitude with the theory prediction (dashed line) based on independently measured system parameters. The inset shows the relevant level diagram and parametric, Bus mediated, photon exchange process. The Bus sidebands are shown on the right for two typical flux modulation amplitudes. \textbf{c}, The measured parametric photon exchange rates, $g_{ij}$, and resulting connectivity graph programmed by the Bus flux modulation. \textbf{d}, The pulse sequence and resulting programmed single photon hopping from Q$_1$ $\rightarrow$ Q$_4$ $\rightarrow$ Q$_7$ $\rightarrow$ Q$_1$ as indicated by the directed lines in \textbf{c}.} 
    \label{fig:q14exchange}
\end{figure*}

We observe the expected parametric coherent photon exchange between pairs of qubits resulting in the characteristic chevron pattern in Fig.~\ref{fig:q14exchange}\textbf{a} for Q$_{1}$ and Q$_{4}$. By repeating the time-domain experiment between Q$_{1}$ and Q$_{4}$ for different Bus flux modulation amplitudes, we extract the photon exchange rate, $g_{14}$, as a function of Bus modulation amplitude. This is shown in Fig.~\ref{fig:q14exchange}\textbf{b} where we overlay the theory prediction based on independently measured system parameters from previous data fits of the Bus spectrum and qubit-Bus coupling. The close alignment between theory prediction and independent experimental observation highlights the accuracy of the novel parametric dynamics model developed in this work. We further characterized the readout and coherence limited long-range fSim gate process fidelity between the two qubits as $\mathcal{F}=68.8 \pm 1.6\%$ (see SI Note~\ref{sec:2qgatecal} for details). 

Similar pairwise coherent photon exchange experiments are carried out between all 5 operational qubits and we identified 8, of a possible 10, pairwise parametric photon exchange routes enabled on-demand by choosing the appropriate Bus flux modulation frequency. The available connections and the maximum observed photon exchange rate for each pair are show in Fig.~\ref{fig:q14exchange}\textbf{c}. Furthermore, we demonstrate programmable and dynamic routing of a photon throughout the network. In Fig.~\ref{fig:q14exchange}\textbf{d} a photon is exchanged on demand among spatially separated qubits Q$_{1}$, Q$_{4}$, and Q$_{7}$. It should also be noted that a long distance coherent photon exchange was observed between Q$_{1}$ and Q$_{8}$ separated by $\sim 9.2~\text{cm}$ along the meandered Bus CPW. The direct photon exchange between qubits mediated by an almost $10~\text{cm}$ CPW demonstrates the possibility of applying this architecture to entangling qubits across standard superconducting quantum chips even as their size scales up in the future.

\section*{Summary and outlook
\label{sec:conlusion}}
In this work, we propose and demonstrate periodic modulation of a multi-mode linear resonator's vacuum electromagnetic field to induce parametric pairwise photon exchange between qubits non-locally. Applying this to a multi-qubit superconducting quantum processor we create an on-chip, reconfigurable, highly connected photonic network \cite{zhang_metamaterial}. We develop a robust and novel theory \cite{WustmannParametricResonance,WallquistSelectiveCoupling} for describing the dynamical behavior of the distributed element resonator under boundary parametric modulation and thoroughly verify that the model achieves good agreement with experimental results. This theoretical framework should be readily applicable to other superconducting microwave circuit systems with parametrically modulated lumped or distributed elements. 

We embedded the novel photon-exchanging framework in an eight-qubit quantum processor. We realize a programmable high connectivity coupling graph between qubits in a subsystem of our device by parametrically inducing photon exchange between qubits separated by up to $\sim9.2$~cm. We find the process fidelity for two-qubit gates based on this photon exchange interaction is likely limited by engineering imperfections such as non-ideal readout, relatively low qubit coherence, and low-frequency noise in flux lines; addressing these straightforward engineering challenges should yield significant increases in the device coherence and process fidelities.

This photon exchange network architecture can serve as a common platform for a variety of novel quantum experiments requiring non-local connectivity or higher dimensional arrays of qubits. Future applications in the exploration of novel many-body physics \cite{Islam2013,Landig2016,Ebadi2021,Joshi2022}, hybrid Boson-spin systems \cite{Puri2019}, or modular quantum computing using entangled modes between two devices \cite{Narla2016,Zhong2021} are all possible.

\section*{Acknowledgements}
We are grateful to L. Chen and R. Naik for conversations and insights. This work was supported by the Quantum Testbed Program of the Advanced Scientific Computing Research for Basic Energy Sciences program, Office of Science of the U.S. Department of Energy under Contract No. DE-AC02-05CH11231.

\newpage
\bibliographystyle{naturemag}
\bibliography{reference/references_jl.bib,reference/literature_list_brian.bib}

\begin{thebibliography}{10}
\expandafter\ifx\csname url\endcsname\relax
  \def\url#1{\texttt{#1}}\fi
\expandafter\ifx\csname urlprefix\endcsname\relax\def\urlprefix{URL }\fi
\providecommand{\bibinfo}[2]{#2}
\providecommand{\eprint}[2][]{\url{#2}}

\bibitem{2019GoogleSupremacy}
\bibinfo{author}{Arute, F.} \emph{et~al.}
\newblock \bibinfo{title}{Quantum supremacy using a programmable
  superconducting processor}.
\newblock \emph{\bibinfo{journal}{Nature}} \textbf{\bibinfo{volume}{574}},
  \bibinfo{pages}{505–510} (\bibinfo{year}{2019}).
\newblock \urlprefix\url{https://www.nature.com/articles/s41586-019-1666-5}.

\bibitem{2020GoogleEnergies}
\bibinfo{author}{Neill, C.} \emph{et~al.}
\newblock \bibinfo{title}{Accurately computing the electronic properties of a
  quantum ring}.
\newblock \emph{\bibinfo{journal}{Nature}} \textbf{\bibinfo{volume}{594}},
  \bibinfo{pages}{508--512} (\bibinfo{year}{2021}).
\newblock \urlprefix\url{https://doi.org/10.1038/s41586-021-03576-2}.

\bibitem{2021GoogleQAOA}
\bibinfo{author}{Harrigan, M.} \emph{et~al.}
\newblock \bibinfo{title}{Quantum approximate optimization of non-planar graph
  problems on a planar superconducting processor}.
\newblock \emph{\bibinfo{journal}{Nature Physics}}  (\bibinfo{year}{2021}).
\newblock \urlprefix\url{https://www.nature.com/articles/s41567-020-01105-y}.

\bibitem{fedorov2022}
\bibinfo{author}{Fedorov, A.~K.}, \bibinfo{author}{Gisin, N.},
  \bibinfo{author}{Beloussov, S.~M.} \& \bibinfo{author}{Lvovsky, A.~I.}
\newblock \bibinfo{title}{Quantum computing at the quantum advantage threshold:
  a down-to-business review}  (\bibinfo{year}{2022}).
\newblock \urlprefix\url{https://arxiv.org/abs/2203.17181}.

\bibitem{OMalley2016}
\bibinfo{author}{O'Malley, P. J.~J.} \emph{et~al.}
\newblock \bibinfo{title}{Scalable quantum simulation of molecular energies}.
\newblock \emph{\bibinfo{journal}{Phys. Rev. X}} \textbf{\bibinfo{volume}{6}},
  \bibinfo{pages}{031007} (\bibinfo{year}{2016}).

\bibitem{Islam2013}
\bibinfo{author}{Islam, R.} \emph{et~al.}
\newblock \bibinfo{title}{Emergence and frustration of magnetism with
  variable-range interactions in a quantum simulator}.
\newblock \emph{\bibinfo{journal}{Science}} \textbf{\bibinfo{volume}{340}},
  \bibinfo{pages}{583--587} (\bibinfo{year}{2013}).
\newblock
  \urlprefix\url{https://www.science.org/doi/abs/10.1126/science.1232296}.
\newblock \eprint{https://www.science.org/doi/pdf/10.1126/science.1232296}.

\bibitem{seetharam2021}
\bibinfo{author}{Seetharam, K.} \emph{et~al.}
\newblock \bibinfo{title}{Digital quantum simulation of nmr experiments}
  (\bibinfo{year}{2021}).
\newblock \urlprefix\url{https://arxiv.org/abs/2109.13298}.

\bibitem{johri2021}
\bibinfo{author}{Johri, S.} \emph{et~al.}
\newblock \bibinfo{title}{Nearest centroid classification on a trapped ion
  quantum computer}.
\newblock \emph{\bibinfo{journal}{npj Quantum Information}}
  \textbf{\bibinfo{volume}{7}}, \bibinfo{pages}{122} (\bibinfo{year}{2021}).
\newblock \urlprefix\url{https://doi.org/10.1038/s41534-021-00456-5}.

\bibitem{Nam2020}
\bibinfo{author}{Nam, Y.} \emph{et~al.}
\newblock \bibinfo{title}{Ground-state energy estimation of the water molecule
  on a trapped-ion quantum computer}.
\newblock \emph{\bibinfo{journal}{npj Quantum Information}}
  \textbf{\bibinfo{volume}{6}}, \bibinfo{pages}{33} (\bibinfo{year}{2020}).
\newblock \urlprefix\url{https://doi.org/10.1038/s41534-020-0259-3}.

\bibitem{Gill2020}
\bibinfo{author}{Gill, S.~S.} \emph{et~al.}
\newblock \bibinfo{title}{Quantum computing: A taxonomy, systematic review and
  future directions}  (\bibinfo{year}{2020}).
\newblock \urlprefix\url{https://arxiv.org/abs/2010.15559}.

\bibitem{Kivlichan2018}
\bibinfo{author}{Kivlichan, I.~D.} \emph{et~al.}
\newblock \bibinfo{title}{Quantum simulation of electronic structure with
  linear depth and connectivity}.
\newblock \emph{\bibinfo{journal}{Phys. Rev. Lett.}}
  \textbf{\bibinfo{volume}{120}}, \bibinfo{pages}{110501}
  (\bibinfo{year}{2018}).
\newblock
  \urlprefix\url{https://link.aps.org/doi/10.1103/PhysRevLett.120.110501}.

\bibitem{Koch2020}
\bibinfo{author}{Koch, D.}, \bibinfo{author}{Martin, B.},
  \bibinfo{author}{Patel, S.}, \bibinfo{author}{Wessing, L.} \&
  \bibinfo{author}{Alsing, P.~M.}
\newblock \bibinfo{title}{Demonstrating nisq era challenges in algorithm design
  on ibm’s 20 qubit quantum computer}.
\newblock \emph{\bibinfo{journal}{AIP Advances}} \textbf{\bibinfo{volume}{10}},
  \bibinfo{pages}{095101} (\bibinfo{year}{2020}).
\newblock \urlprefix\url{https://doi.org/10.1063/5.0015526}.

\bibitem{svore2006}
\bibinfo{author}{Svore, K.~M.}, \bibinfo{author}{DiVincenzo, D.~P.} \&
  \bibinfo{author}{Terhal, B.~M.}
\newblock \bibinfo{title}{Noise threshold for a fault-tolerant two-dimensional
  lattice architecture}  (\bibinfo{year}{2006}).
\newblock \urlprefix\url{https://arxiv.org/abs/quant-ph/0604090}.

\bibitem{Korenblit2012}
\bibinfo{author}{Korenblit, S.} \emph{et~al.}
\newblock \bibinfo{title}{Quantum simulation of spin models on an arbitrary
  lattice with trapped ions}.
\newblock \emph{\bibinfo{journal}{New Journal of Physics}}
  \textbf{\bibinfo{volume}{14}}, \bibinfo{pages}{095024}
  (\bibinfo{year}{2012}).
\newblock \urlprefix\url{https://doi.org/10.1088/1367-2630/14/9/095024}.

\bibitem{Vaidya2018}
\bibinfo{author}{Vaidya, V.~D.} \emph{et~al.}
\newblock \bibinfo{title}{Tunable-range, photon-mediated atomic interactions in
  multimode cavity qed}.
\newblock \emph{\bibinfo{journal}{Phys. Rev. X}} \textbf{\bibinfo{volume}{8}},
  \bibinfo{pages}{011002} (\bibinfo{year}{2018}).
\newblock \urlprefix\url{https://link.aps.org/doi/10.1103/PhysRevX.8.011002}.

\bibitem{Wright2019}
\bibinfo{author}{Wright, K.} \emph{et~al.}
\newblock \bibinfo{title}{Benchmarking an 11-qubit quantum computer}.
\newblock \emph{\bibinfo{journal}{Nature Communications}}
  \textbf{\bibinfo{volume}{10}}, \bibinfo{pages}{5464} (\bibinfo{year}{2019}).
\newblock \urlprefix\url{https://doi.org/10.1038/s41467-019-13534-2}.

\bibitem{Manovitz2020}
\bibinfo{author}{Manovitz, T.}, \bibinfo{author}{Shapira, Y.},
  \bibinfo{author}{Akerman, N.}, \bibinfo{author}{Stern, A.} \&
  \bibinfo{author}{Ozeri, R.}
\newblock \bibinfo{title}{Quantum simulations with complex geometries and
  synthetic gauge fields in a trapped ion chain}.
\newblock \emph{\bibinfo{journal}{PRX Quantum}} \textbf{\bibinfo{volume}{1}},
  \bibinfo{pages}{020303} (\bibinfo{year}{2020}).
\newblock \urlprefix\url{https://link.aps.org/doi/10.1103/PRXQuantum.1.020303}.

\bibitem{Periwal2021}
\bibinfo{author}{Periwal, A.} \emph{et~al.}
\newblock \bibinfo{title}{Programmable interactions and emergent geometry in an
  array of atom clouds}.
\newblock \emph{\bibinfo{journal}{Nature}} \textbf{\bibinfo{volume}{600}},
  \bibinfo{pages}{630--635} (\bibinfo{year}{2021}).
\newblock \urlprefix\url{https://doi.org/10.1038/s41586-021-04156-0}.

\bibitem{Bluvstein2022}
\bibinfo{author}{Bluvstein, D.} \emph{et~al.}
\newblock \bibinfo{title}{A quantum processor based on coherent transport of
  entangled atom arrays}.
\newblock \emph{\bibinfo{journal}{Nature}} \textbf{\bibinfo{volume}{604}},
  \bibinfo{pages}{451--456} (\bibinfo{year}{2022}).
\newblock \urlprefix\url{https://doi.org/10.1038%2Fs41586-022-04592-6}.

\bibitem{Hamerly2019}
\bibinfo{author}{Hamerly, R.} \emph{et~al.}
\newblock \bibinfo{title}{Experimental investigation of performance differences
  between coherent ising machines and a quantum annealer}.
\newblock \emph{\bibinfo{journal}{Science Advances}}
  \textbf{\bibinfo{volume}{5}} (\bibinfo{year}{2019}).
\newblock \urlprefix\url{https://doi.org/10.1126%2Fsciadv.aau0823}.

\bibitem{Altman2021}
\bibinfo{author}{Altman, E.} \emph{et~al.}
\newblock \bibinfo{title}{Quantum simulators: Architectures and opportunities}.
\newblock \emph{\bibinfo{journal}{PRX Quantum}} \textbf{\bibinfo{volume}{2}},
  \bibinfo{pages}{017003} (\bibinfo{year}{2021}).
\newblock \urlprefix\url{https://link.aps.org/doi/10.1103/PRXQuantum.2.017003}.

\bibitem{Nandkishore2015}
\bibinfo{author}{Nandkishore, R.} \& \bibinfo{author}{Huse, D.~A.}
\newblock \bibinfo{title}{Many-body localization and thermalization in quantum
  statistical mechanics}.
\newblock \emph{\bibinfo{journal}{Annual Review of Condensed Matter Physics}}
  \textbf{\bibinfo{volume}{6}}, \bibinfo{pages}{15--38} (\bibinfo{year}{2015}).
\newblock
  \urlprefix\url{https://doi.org/10.1146/annurev-conmatphys-031214-014726}.
\newblock \eprint{https://doi.org/10.1146/annurev-conmatphys-031214-014726}.

\bibitem{Joshi2020}
\bibinfo{author}{Joshi, M.~K.} \emph{et~al.}
\newblock \bibinfo{title}{Quantum information scrambling in a trapped-ion
  quantum simulator with tunable range interactions}.
\newblock \emph{\bibinfo{journal}{Phys. Rev. Lett.}}
  \textbf{\bibinfo{volume}{124}}, \bibinfo{pages}{240505}
  (\bibinfo{year}{2020}).
\newblock
  \urlprefix\url{https://link.aps.org/doi/10.1103/PhysRevLett.124.240505}.

\bibitem{Landig2016}
\bibinfo{author}{Landig, R.} \emph{et~al.}
\newblock \bibinfo{title}{Quantum phases from competing short- and long-range
  interactions in an optical lattice}.
\newblock \emph{\bibinfo{journal}{Nature}} \textbf{\bibinfo{volume}{532}},
  \bibinfo{pages}{476--479} (\bibinfo{year}{2016}).
\newblock \urlprefix\url{https://doi.org/10.1038/nature17409}.

\bibitem{Ebadi2021}
\bibinfo{author}{Ebadi, S.} \emph{et~al.}
\newblock \bibinfo{title}{Quantum phases of matter on a 256-atom programmable
  quantum simulator}.
\newblock \emph{\bibinfo{journal}{Nature}} \textbf{\bibinfo{volume}{595}},
  \bibinfo{pages}{227--232} (\bibinfo{year}{2021}).
\newblock \urlprefix\url{https://doi.org/10.1038/s41586-021-03582-4}.

\bibitem{Joshi2022}
\bibinfo{author}{Joshi, M.~K.} \emph{et~al.}
\newblock \bibinfo{title}{Observing emergent hydrodynamics in a long-range
  quantum magnet}.
\newblock \emph{\bibinfo{journal}{Science}} \textbf{\bibinfo{volume}{376}},
  \bibinfo{pages}{720--724} (\bibinfo{year}{2022}).
\newblock
  \urlprefix\url{https://www.science.org/doi/abs/10.1126/science.abk2400}.
\newblock \eprint{https://www.science.org/doi/pdf/10.1126/science.abk2400}.

\bibitem{Jurcevic_2021}
\bibinfo{author}{Jurcevic, P.} \emph{et~al.}
\newblock \bibinfo{title}{Demonstration of quantum volume 64 on a
  superconducting quantum computing system}.
\newblock \emph{\bibinfo{journal}{Quantum Science and Technology}}
  \textbf{\bibinfo{volume}{6}}, \bibinfo{pages}{025020} (\bibinfo{year}{2021}).
\newblock \urlprefix\url{https://doi.org/10.1088/2058-9565/abe519}.

\bibitem{Zhang2017}
\bibinfo{author}{Zhang, J.} \emph{et~al.}
\newblock \bibinfo{title}{Observation of a discrete time crystal}.
\newblock \emph{\bibinfo{journal}{Nature}} \textbf{\bibinfo{volume}{543}},
  \bibinfo{pages}{217--220} (\bibinfo{year}{2017}).
\newblock \urlprefix\url{https://doi.org/10.1038/nature21413}.

\bibitem{Landsman2019}
\bibinfo{author}{Landsman, K.~A.} \emph{et~al.}
\newblock \bibinfo{title}{Verified quantum information scrambling}.
\newblock \emph{\bibinfo{journal}{Nature}} \textbf{\bibinfo{volume}{567}},
  \bibinfo{pages}{61--65} (\bibinfo{year}{2019}).
\newblock \urlprefix\url{https://doi.org/10.1038/s41586-019-0952-6}.

\bibitem{Lu_MSGate}
\bibinfo{author}{Lu, M.} \emph{et~al.}
\newblock \bibinfo{title}{Multipartite entanglement in rabi driven
  superconducting qubits} (\bibinfo{year}{2022}).
\newblock \urlprefix\url{https://arxiv.org/abs/2207.00130}.

\bibitem{zhang_metamaterial}
\bibinfo{author}{Zhang, X.}, \bibinfo{author}{Kim, E.}, \bibinfo{author}{Mark,
  D.~K.}, \bibinfo{author}{Choi, S.} \& \bibinfo{author}{Painter, O.}
\newblock \bibinfo{title}{A superconducting quantum simulator based on a
  photonic-bandgap metamaterial}.
\newblock \emph{\bibinfo{journal}{Science}} \textbf{\bibinfo{volume}{379}},
  \bibinfo{pages}{278--283} (\bibinfo{year}{2023}).
\newblock
  \urlprefix\url{https://www.science.org/doi/abs/10.1126/science.ade7651}.
\newblock \eprint{https://www.science.org/doi/pdf/10.1126/science.ade7651}.

\bibitem{NoriFluxQubitBus}
\bibinfo{author}{Stassi, R.}, \bibinfo{author}{Cirio, M.} \&
  \bibinfo{author}{Nori, F.}
\newblock \bibinfo{title}{Scalable quantum computer with superconducting
  circuits in the ultrastrong coupling regime}.
\newblock \emph{\bibinfo{journal}{npj Quantum Information}}
  \textbf{\bibinfo{volume}{6}}, \bibinfo{pages}{67} (\bibinfo{year}{2020}).
\newblock \urlprefix\url{https://doi.org/10.1038/s41534-020-00294-x}.

\bibitem{Onodera2020}
\bibinfo{author}{Onodera, T.}, \bibinfo{author}{Ng, E.} \&
  \bibinfo{author}{McMahon, P.~L.}
\newblock \bibinfo{title}{A quantum annealer with fully programmable all-to-all
  coupling via floquet engineering}.
\newblock \emph{\bibinfo{journal}{npj Quantum Information}}
  \textbf{\bibinfo{volume}{6}}, \bibinfo{pages}{48} (\bibinfo{year}{2020}).
\newblock \urlprefix\url{https://doi.org/10.1038/s41534-020-0279-z}.

\bibitem{Rosenberg20173D}
\bibinfo{author}{Rosenberg, D.} \emph{et~al.}
\newblock \bibinfo{title}{3d integrated superconducting qubits}.
\newblock \emph{\bibinfo{journal}{npj Quantum Information}}
  \textbf{\bibinfo{volume}{3}}, \bibinfo{pages}{42} (\bibinfo{year}{2017}).
\newblock \urlprefix\url{https://doi.org/10.1038/s41534-017-0044-0}.

\bibitem{Yost2020}
\bibinfo{author}{Yost, D. R.~W.} \emph{et~al.}
\newblock \bibinfo{title}{Solid-state qubits integrated with superconducting
  through-silicon vias}.
\newblock \emph{\bibinfo{journal}{npj Quantum Information}}
  \textbf{\bibinfo{volume}{6}}, \bibinfo{pages}{59} (\bibinfo{year}{2020}).
\newblock \urlprefix\url{https://doi.org/10.1038/s41534-020-00289-8}.

\bibitem{ChowCR}
\bibinfo{author}{Chow, J.~M.} \emph{et~al.}
\newblock \bibinfo{title}{Simple all-microwave entangling gate for
  fixed-frequency superconducting qubits}.
\newblock \emph{\bibinfo{journal}{Phys. Rev. Lett.}}
  \textbf{\bibinfo{volume}{107}}, \bibinfo{pages}{080502}
  (\bibinfo{year}{2011}).
\newblock
  \urlprefix\url{https://link.aps.org/doi/10.1103/PhysRevLett.107.080502}.

\bibitem{Paik_RIPGate}
\bibinfo{author}{Paik, H.} \emph{et~al.}
\newblock \bibinfo{title}{Experimental demonstration of a resonator-induced
  phase gate in a multiqubit circuit-qed system}.
\newblock \emph{\bibinfo{journal}{Phys. Rev. Lett.}}
  \textbf{\bibinfo{volume}{117}}, \bibinfo{pages}{250502}
  (\bibinfo{year}{2016}).
\newblock
  \urlprefix\url{https://link.aps.org/doi/10.1103/PhysRevLett.117.250502}.

\bibitem{CasparisVoltageTunableBus}
\bibinfo{author}{Casparis, L.} \emph{et~al.}
\newblock \bibinfo{title}{Voltage-controlled superconducting quantum bus}.
\newblock \emph{\bibinfo{journal}{Phys. Rev. B}} \textbf{\bibinfo{volume}{99}},
  \bibinfo{pages}{085434} (\bibinfo{year}{2019}).
\newblock \urlprefix\url{https://link.aps.org/doi/10.1103/PhysRevB.99.085434}.

\bibitem{Sandberg2008}
\bibinfo{author}{Sandberg, M.} \emph{et~al.}
\newblock \bibinfo{title}{Tuning the field in a microwave resonator faster than
  the photon lifetime}.
\newblock \emph{\bibinfo{journal}{Applied Physics Letters}}
  \textbf{\bibinfo{volume}{92}}, \bibinfo{pages}{203501}
  (\bibinfo{year}{2008}).
\newblock \urlprefix\url{https://doi.org/10.1063/1.2929367}.
\newblock \eprint{https://doi.org/10.1063/1.2929367}.

\bibitem{Palacios-Laloy2008}
\bibinfo{author}{Palacios-Laloy, A.} \emph{et~al.}
\newblock \bibinfo{title}{Tunable resonators for quantum circuits}.
\newblock \emph{\bibinfo{journal}{Journal of Low Temperature Physics}}
  \textbf{\bibinfo{volume}{151}}, \bibinfo{pages}{1034--1042}
  (\bibinfo{year}{2008}).
\newblock \urlprefix\url{https://doi.org/10.1007/s10909-008-9774-x}.

\bibitem{Castellanos-Beltran2007}
\bibinfo{author}{Castellanos-Beltran, M.~A.} \& \bibinfo{author}{Lehnert,
  K.~W.}
\newblock \bibinfo{title}{Widely tunable parametric amplifier based on a
  superconducting quantum interference device array resonator}.
\newblock \emph{\bibinfo{journal}{Applied Physics Letters}}
  \textbf{\bibinfo{volume}{91}}, \bibinfo{pages}{083509}
  (\bibinfo{year}{2007}).
\newblock \urlprefix\url{https://doi.org/10.1063/1.2773988}.
\newblock \eprint{https://doi.org/10.1063/1.2773988}.

\bibitem{JohanssonCasimir1}
\bibinfo{author}{Johansson, J.~R.}, \bibinfo{author}{Johansson, G.},
  \bibinfo{author}{Wilson, C.~M.} \& \bibinfo{author}{Nori, F.}
\newblock \bibinfo{title}{Dynamical casimir effect in a superconducting
  coplanar waveguide}.
\newblock \emph{\bibinfo{journal}{Phys. Rev. Lett.}}
  \textbf{\bibinfo{volume}{103}}, \bibinfo{pages}{147003}
  (\bibinfo{year}{2009}).
\newblock
  \urlprefix\url{https://link.aps.org/doi/10.1103/PhysRevLett.103.147003}.

\bibitem{JohanssonCasimir2}
\bibinfo{author}{Johansson, J.~R.}, \bibinfo{author}{Johansson, G.},
  \bibinfo{author}{Wilson, C.~M.} \& \bibinfo{author}{Nori, F.}
\newblock \bibinfo{title}{Dynamical casimir effect in superconducting microwave
  circuits}.
\newblock \emph{\bibinfo{journal}{Phys. Rev. A}} \textbf{\bibinfo{volume}{82}},
  \bibinfo{pages}{052509} (\bibinfo{year}{2010}).
\newblock \urlprefix\url{https://link.aps.org/doi/10.1103/PhysRevA.82.052509}.

\bibitem{WustmannParametricResonance}
\bibinfo{author}{Wustmann, W.} \& \bibinfo{author}{Shumeiko, V.}
\newblock \bibinfo{title}{Parametric resonance in tunable superconducting
  cavities}.
\newblock \emph{\bibinfo{journal}{Phys. Rev. B}} \textbf{\bibinfo{volume}{87}},
  \bibinfo{pages}{184501} (\bibinfo{year}{2013}).
\newblock \urlprefix\url{https://link.aps.org/doi/10.1103/PhysRevB.87.184501}.

\bibitem{WallquistSelectiveCoupling}
\bibinfo{author}{Wallquist, M.}, \bibinfo{author}{Shumeiko, V.~S.} \&
  \bibinfo{author}{Wendin, G.}
\newblock \bibinfo{title}{Selective coupling of superconducting charge qubits
  mediated by a tunable stripline cavity}.
\newblock \emph{\bibinfo{journal}{Phys. Rev. B}} \textbf{\bibinfo{volume}{74}},
  \bibinfo{pages}{224506} (\bibinfo{year}{2006}).
\newblock \urlprefix\url{https://link.aps.org/doi/10.1103/PhysRevB.74.224506}.

\bibitem{Puri2019}
\bibinfo{author}{Puri, S.} \emph{et~al.}
\newblock \bibinfo{title}{Stabilized cat in a driven nonlinear cavity: A
  fault-tolerant error syndrome detector}.
\newblock \emph{\bibinfo{journal}{PRX}} \textbf{\bibinfo{volume}{9}},
  \bibinfo{pages}{041009} (\bibinfo{year}{2019}).
\newblock \urlprefix\url{https://link.aps.org/doi/10.1103/PhysRevX.9.041009}.

\bibitem{Narla2016}
\bibinfo{author}{Narla, A.} \emph{et~al.}
\newblock \bibinfo{title}{Robust concurrent remote entanglement between two
  superconducting qubits}.
\newblock \emph{\bibinfo{journal}{PRX}} \textbf{\bibinfo{volume}{6}},
  \bibinfo{pages}{031036} (\bibinfo{year}{2016}).
\newblock \urlprefix\url{https://link.aps.org/doi/10.1103/PhysRevX.6.031036}.

\bibitem{Zhong2021}
\bibinfo{author}{Zhong, Y.} \emph{et~al.}
\newblock \bibinfo{title}{Deterministic multi-qubit entanglement in a quantum
  network}.
\newblock \emph{\bibinfo{journal}{Nature}} \textbf{\bibinfo{volume}{590}},
  \bibinfo{pages}{571--575} (\bibinfo{year}{2021}).
\newblock \urlprefix\url{https://doi.org/10.1038/s41586-021-03288-7}.

\bibitem{Kosen_2022}
\bibinfo{author}{Kosen, S.} \emph{et~al.}
\newblock \bibinfo{title}{Building blocks of a flip-chip integrated
  superconducting quantum processor}.
\newblock \emph{\bibinfo{journal}{Quantum Science and Technology}}
  \textbf{\bibinfo{volume}{7}}, \bibinfo{pages}{035018} (\bibinfo{year}{2022}).
\newblock \urlprefix\url{https://doi.org/10.1088/2058-9565/ac734b}.

\bibitem{Blais2020Review}
\bibinfo{author}{Blais, A.}, \bibinfo{author}{Grimsmo, A.~L.},
  \bibinfo{author}{Girvin, S.~M.} \& \bibinfo{author}{Wallraff, A.}
\newblock \bibinfo{title}{Circuit quantum electrodynamics}.
\newblock \emph{\bibinfo{journal}{Rev. Mod. Phys.}}
  \textbf{\bibinfo{volume}{93}}, \bibinfo{pages}{025005}
  (\bibinfo{year}{2021}).
\newblock
  \urlprefix\url{https://link.aps.org/doi/10.1103/RevModPhys.93.025005}.

\bibitem{Foxen2020fSim}
\bibinfo{author}{Foxen, B.} \emph{et~al.}
\newblock \bibinfo{title}{Demonstrating a continuous set of two-qubit gates for
  near-term quantum algorithms}.
\newblock \emph{\bibinfo{journal}{Phys. Rev. Lett.}}
  \textbf{\bibinfo{volume}{125}}, \bibinfo{pages}{120504}
  (\bibinfo{year}{2020}).
\newblock
  \urlprefix\url{https://link.aps.org/doi/10.1103/PhysRevLett.125.120504}.

\bibitem{Abrams2020}
\bibinfo{author}{Abrams, D.~M.}, \bibinfo{author}{Didier, N.},
  \bibinfo{author}{Johnson, B.~R.}, \bibinfo{author}{Silva, M. P.~d.} \&
  \bibinfo{author}{Ryan, C.~A.}
\newblock \bibinfo{title}{Implementation of xy entangling gates with a single
  calibrated pulse}.
\newblock \emph{\bibinfo{journal}{Nature Electronics}}
  \textbf{\bibinfo{volume}{3}}, \bibinfo{pages}{744--750}
  (\bibinfo{year}{2020}).
\newblock \urlprefix\url{https://doi.org/10.1038/s41928-020-00498-1}.

\bibitem{OBrien_QPT}
\bibinfo{author}{O'Brien, J.~L.} \emph{et~al.}
\newblock \bibinfo{title}{Quantum process tomography of a controlled-not gate}.
\newblock \emph{\bibinfo{journal}{Phys. Rev. Lett.}}
  \textbf{\bibinfo{volume}{93}}, \bibinfo{pages}{080502}
  (\bibinfo{year}{2004}).
\newblock
  \urlprefix\url{https://link.aps.org/doi/10.1103/PhysRevLett.93.080502}.

\bibitem{Sato2012MultiModePhotonics}
\bibinfo{author}{Sato, Y.} \emph{et~al.}
\newblock \bibinfo{title}{Strong coupling between distant photonic nanocavities
  and its dynamic control}.
\newblock \emph{\bibinfo{journal}{Nature Photonics}}
  \textbf{\bibinfo{volume}{6}}, \bibinfo{pages}{56--61} (\bibinfo{year}{2012}).
\newblock \urlprefix\url{https://doi.org/10.1038/nphoton.2011.286}.

\bibitem{Dawkins_CohLimit}
\bibinfo{author}{Dawkins, H.}, \bibinfo{author}{Wallman, J.} \&
  \bibinfo{author}{Emerson, J.}
\newblock \bibinfo{title}{Combining ${T}_{1}$ and ${T}_{2}$ estimation with
  randomized benchmarking and bounding the diamond distance}.
\newblock \emph{\bibinfo{journal}{Phys. Rev. A}}
  \textbf{\bibinfo{volume}{102}}, \bibinfo{pages}{022220}
  (\bibinfo{year}{2020}).
\newblock \urlprefix\url{https://link.aps.org/doi/10.1103/PhysRevA.102.022220}.

\bibitem{GARBOWJazz2}
\bibinfo{author}{Garbow, J.}, \bibinfo{author}{Weitekamp, D.} \&
  \bibinfo{author}{Pines, A.}
\newblock \bibinfo{title}{Bilinear rotation decoupling of homonuclear scalar
  interactions}.
\newblock \emph{\bibinfo{journal}{Chemical Physics Letters}}
  \textbf{\bibinfo{volume}{93}}, \bibinfo{pages}{504--509}
  (\bibinfo{year}{1982}).
\newblock
  \urlprefix\url{https://www.sciencedirect.com/science/article/pii/0009261482832296}.

\bibitem{TakitaJazz1}
\bibinfo{author}{Takita, M.}, \bibinfo{author}{Cross, A.~W.},
  \bibinfo{author}{C\'orcoles, A.~D.}, \bibinfo{author}{Chow, J.~M.} \&
  \bibinfo{author}{Gambetta, J.~M.}
\newblock \bibinfo{title}{Experimental demonstration of fault-tolerant state
  preparation with superconducting qubits}.
\newblock \emph{\bibinfo{journal}{Phys. Rev. Lett.}}
  \textbf{\bibinfo{volume}{119}}, \bibinfo{pages}{180501}
  (\bibinfo{year}{2017}).
\newblock
  \urlprefix\url{https://link.aps.org/doi/10.1103/PhysRevLett.119.180501}.

\bibitem{BlaisCQED2004}
\bibinfo{author}{Blais, A.}, \bibinfo{author}{Huang, R.-S.},
  \bibinfo{author}{Wallraff, A.}, \bibinfo{author}{Girvin, S.~M.} \&
  \bibinfo{author}{Schoelkopf, R.~J.}
\newblock \bibinfo{title}{Cavity quantum electrodynamics for superconducting
  electrical circuits: An architecture for quantum computation}.
\newblock \emph{\bibinfo{journal}{Phys. Rev. A}} \textbf{\bibinfo{volume}{69}},
  \bibinfo{pages}{062320} (\bibinfo{year}{2004}).
\newblock \urlprefix\url{https://link.aps.org/doi/10.1103/PhysRevA.69.062320}.

\bibitem{Kreikebaum_2020}
\bibinfo{author}{Kreikebaum, J.~M.}, \bibinfo{author}{O'Brien, K.~P.},
  \bibinfo{author}{Morvan, A.} \& \bibinfo{author}{Siddiqi, I.}
\newblock \bibinfo{title}{Improving wafer-scale josephson junction resistance
  variation in superconducting quantum coherent circuits}.
\newblock \emph{\bibinfo{journal}{Superconductor Science and Technology}}
  \textbf{\bibinfo{volume}{33}}, \bibinfo{pages}{06LT02}
  (\bibinfo{year}{2020}).
\newblock \urlprefix\url{https://doi.org/10.1088/1361-6668/ab8617}.

\bibitem{morvan2021optimizing}
\bibinfo{author}{Morvan, A.}, \bibinfo{author}{Chen, L.},
  \bibinfo{author}{Larson, J.~M.}, \bibinfo{author}{Santiago, D.~I.} \&
  \bibinfo{author}{Siddiqi, I.}
\newblock \bibinfo{title}{Optimizing frequency allocation for fixed-frequency
  superconducting quantum processors} (\bibinfo{year}{2021}).
\newblock \eprint{2112.01634}.

\bibitem{ZhangLaser2022}
\bibinfo{author}{Zhang, E.~J.} \emph{et~al.}
\newblock \bibinfo{title}{High-performance superconducting quantum processors
  via laser annealing of transmon qubits}.
\newblock \emph{\bibinfo{journal}{Science Advances}}
  \textbf{\bibinfo{volume}{8}}, \bibinfo{pages}{eabi6690}
  (\bibinfo{year}{2022}).
\newblock
  \urlprefix\url{https://www.science.org/doi/abs/10.1126/sciadv.abi6690}.
\newblock \eprint{https://www.science.org/doi/pdf/10.1126/sciadv.abi6690}.

\bibitem{Hertzberg2021}
\bibinfo{author}{Hertzberg, J.~B.} \emph{et~al.}
\newblock \bibinfo{title}{Laser-annealing josephson junctions for yielding
  scaled-up superconducting quantum processors}.
\newblock \emph{\bibinfo{journal}{npj Quantum Information}}
  \textbf{\bibinfo{volume}{7}}, \bibinfo{pages}{129} (\bibinfo{year}{2021}).
\newblock \urlprefix\url{https://doi.org/10.1038/s41534-021-00464-5}.

\bibitem{KimLaser2022}
\bibinfo{author}{Kim, H.} \emph{et~al.}
\newblock \bibinfo{title}{Effects of laser-annealing on fixed-frequency
  superconducting qubits} (\bibinfo{year}{2022}).
\newblock \urlprefix\url{https://arxiv.org/abs/2206.03099}.

\bibitem{Hutchings2017}
\bibinfo{author}{Hutchings, M.~D.} \emph{et~al.}
\newblock \bibinfo{title}{Tunable superconducting qubits with flux-independent
  coherence}.
\newblock \emph{\bibinfo{journal}{Phys. Rev. Applied}}
  \textbf{\bibinfo{volume}{8}}, \bibinfo{pages}{044003} (\bibinfo{year}{2017}).
\newblock
  \urlprefix\url{https://link.aps.org/doi/10.1103/PhysRevApplied.8.044003}.

\end{thebibliography}

\newpage 
\clearpage
\pagebreak

\onecolumngrid
\begin{center}
\textbf{\large Supplementary Information for: ``Demonstration of Dynamically Reconfigurable Long-Range Photon Exchange in a Multi-Qubit Superconducting Quantum Processor''}
\end{center}
\twocolumngrid

\setcounter{equation}{0}
\setcounter{figure}{0}
\setcounter{table}{0}
\setcounter{page}{1}
\renewcommand{\thefigure}{S\arabic{figure}}
\renewcommand{\theequation}{S\arabic{equation}}
\makeatletter

\section{Derivation of Bus Equations of Motion \label{sec:busEOM}}
The Bus bulk and boundary phase fields satisfy Eqn.~\ref{eqn:eombulk} and Eqn.~\ref{eqn:eombndry} respectively. Here we derive these equations of motion (EOMs) starting from a lumped LC model of a transmission line with boundary SQUIDs. The derivation follows closely previous work in \cite{WustmannParametricResonance} and \cite{WallquistSelectiveCoupling} but is generalized to the case where both ends of the Bus are terminated by SQUIDs. 

A circuit model of the tunable Bus is shown in Fig.~\ref{fig:spidernetcircuit}\textbf{b}. The bulk CPW is modeled as a series of $N$ inductors, $L$, in series, shunted to ground by capacitors, $C$. We define $N$ nodes with node fluxes $\phi_{1},\dots,\phi_{N}$. At the two boundaries of the Bus are SQUIDs composed of junctions with Josephson energies $E_{\text{J},sj}$. Then $s=\pm$ denotes the SQUID at $x=s \ell/2$ and $j=1,2$ labels the junctions within SQUID-$s$. The junctions within the SQUIDs have node fluxes $\phi_{sj}$ and finite capacitance $C_{sj}$. With these definitions of the parameters and degrees of freedom (node fluxes) we write down the classical Lagrangian describing the system. For the bulk CPW,
\begin{multline}
    \mathcal{L}_{\text{cav}}=\int dt\left(\frac{\hbar}{2e}\right)^{2}\Bigg\{\sum_{i=1}^{N-1}\left(\frac{C\dot{\phi}_{i}^{2}}{2}-\frac{(\phi_{i+1}-\phi_{i})^{2}}{2L}\right) \\
    +\frac{C\dot{\phi}_{N}^{2}}{2}-\frac{(\phi_{+}-\phi_{N})^{2}}{2L}-\frac{(\phi_{-}-\phi_{1})^{2}}{2L}\Bigg\}
\end{multline}
where $\phi_{s}=(\phi_{s1}+\phi_{s2})/2$ is the node flux for SQUID-$s$. The first three terms give the capacitive (kinetic) and inductive (potential) energies of the $N$ interior/bulk nodes. The last two terms give the inductive energy between the junction nodes and the first and last bulk nodes. Next is the Lagrangian for the SQUIDs,
\begin{equation}
    \mathcal{L}_{\text{SQ}}=\int dt\sum_{s=\pm}\sum_{j=1,2}\left[\left(\frac{\hbar}{2e}\right)^{2}\frac{C_{sj}\dot{\phi}_{sj}^{2}}{2}+E_{\text{J}sj}\cos\phi_{sj}\right].
\end{equation}
The full Lagrangian $\mathcal{L}$ is simply a sum of these parts, $\mathcal{L}=\mathcal{L}_{\text{cav}}+\mathcal{L}_{\text{SQ}}$.

Now we take the continuum limit to find the equations of motion for the phase field, $\phi(x,t)$. We define $C_{0}$ and $L_{0}$ as the capacitance and inductance per unit length along the Bus. Then $C=C_{0}\Delta x$ and $L=L_{0}\Delta x$ where $\Delta x = \ell/N$ is the node separation in the discretized model. The continuum limit involves taking the limits $\Delta x \rightarrow 0$ and $N \rightarrow \infty$ with $N\Delta x=\ell$ held fixed. This turns the sums in $\mathcal{L}$ into Riemann sums and the limit yields an integral. The continuum Lagrangian is
\begin{multline}
    \mathcal{L}=\int dt\Bigg\{\left(\frac{\hbar}{2e}\right)^{2}\int_{-d/2}^{d/2}dx\left[\frac{C_{0}\dot{\phi}^{2}}{2}-\frac{1}{2L_{0}}\left(\frac{d\phi}{dx}\right)^{2}\right] \\
    +\sum_{s=\pm}\left[\left(\frac{\hbar}{2e}\right)^{2}\frac{2C_{s}\dot{\phi}_{s}^{2}}{2}+2E_{\text{J}s}\cos f_{s}\cos\phi_{s}\right]\Bigg\}
\end{multline}
after using $C_{s}=C_{sj}$ for symmetric SQUIDs. Flux quantization requires $\phi_{s1}-\phi_{s2}=f_{s}$ where $f_{s}$ is the external flux threading boundary SQUID-$s$. Now we consider fluctuations of the superconducting phase field $\phi(x,t)=\bar{\phi}(x,t)+\delta\phi(x,t)$, where $\bar{\phi}(x,t)$ is the stationary point of the action and $\delta\phi(x,t)$ describe the fluctuations around this stationary point. We plug this form into the Lagrangian and group terms by their order in $\delta\phi$, up to $\mathcal{O}(\delta \phi)$. For ease of notation we drop the bar over the stationary point solution,
\begin{multline}
    \mathcal{L}=\int dt\Bigg\{\left(\frac{\hbar}{2e}\right)^{2}\int_{-d/2}^{d/2}dx\left[\frac{C_{0}\dot{\phi}^{2}}{2}-\frac{\phi'^{2}}{2L_{0}}\right] \\
    +\sum_{s=\pm}\left[\left(\frac{\hbar}{2e}\right)^{2}\frac{2C_{s}\dot{\phi}_{s}^{2}}{2}+2E_{\text{J}s}\cos\varphi_{s}\cos\phi_{s}\right] \\
    +\left(\frac{\hbar}{2e}\right)^{2}\int_{-d/2}^{d/2}dx\left[C_{0}\dot{\phi}\delta\dot{\phi}-\frac{1}{L_{0}}\phi'\delta\phi'\right] \\
    +\sum_{s=\pm}\left[\left(\frac{\hbar}{2e}\right)^{2}2C_{s}\dot{\phi}_{s}\delta\dot{\phi}_{s}-2E_{\text{J}s}\cos\varphi_{s}\sin\phi_{s}\delta\phi_{s}\right]
\end{multline}
where $\delta \phi_{s}=\delta \phi(s\ell/2,t)$. The zeroth order terms simply give the action of the stationary point solution. Setting the first order variations (in the bulk and at the boundaries) to zero yields the equations of motion for the field. After integrating by parts and setting the first order variations equal to zero we find
\begin{multline}
    0=-\left(\frac{\hbar}{2e}\right)^{2}\int_{-d/2}^{d/2}dx\left[C_{0}\ddot{\phi}-\frac{1}{L_{0}}\phi''\right]\delta\phi \\
    -\sum_{s=\pm}\left[\frac{\ddot{\phi}_{s}}{E_{\text{C}}}+2E_{\text{J}}\cos f_{s}\sin\phi_{s}+sE_{L}d\phi_{s}'\right]\delta\phi_{s}
\end{multline}
where we assume the SQUIDs are identical so $E_{\text{J}+}=E_{\text{J}-}=E_{\text{J}}$ and $C_{+}=C_{-}=C_{\text{J}}$ and make the definitions $E_{\text{C}}=(2e/\hbar)^{2}/2C_{\text{J}}$ and $E_{\mathrm{L}}=(\hbar/2e)^{2}/dL_{0}$ for the SQUID charging and inductive energies, respectively. Setting the coefficient of the bulk field $\delta \phi(x,t)$ to zero in the first term yields the bulk EOM, Eqn.~\ref{eqn:eombulk}, which is simply the wave equation with the speed of light in the Bus CPW $v=1/\sqrt{L_{0}C_{0}}$. Similarly for the coefficient of $\delta \phi_{s}$ we recover Eqn.~\ref{eqn:eombndry} for the boundary phase.

In the case of a time independent or DC flux bias $f_{\pm}(t)=F_{\pm}$ we can substitute an ansatz, $\phi(x,t)=Ae^{i(kx-\omega t)} + Be^{i(kx+\omega t)}$, into the equations of motion. Substituting the ansatz in the bulk equation of motion, Eqn.~\ref{eqn:eombulk}, yields the dispersion relation, $\omega=vk$. Substituting the ansatz in the boundary equations of motion, Eqn.~\ref{eqn:eombndry}, yields a transcendental equation,
\begin{equation}
    \lambda_{k}(F_{-})+\lambda_{k}(F_{+})=\left[\lambda_{k}(F_{-})\lambda_{k}(F_{+})-1\right]\tan{(kd)}
    \label{eqn:buseigenmode}
\end{equation}
where for convenience we define
\begin{equation}
    \lambda_{k}(F)=\frac{1}{E_{L}k\ell}\left[\frac{(kv)^{2}}{E_{\mathrm{C}}}-2E_{\mathrm{J}}\cos F\right].
\end{equation}
Solutions to this equation for $k$ give the eigenfrequencies of the tunable resonator, $\omega=vk$, and can be obtained numerically. The dashed curves in Fig.~\ref{fig:spidernetcircuit}\textbf{b} show Bus mode frequencies extracted in this way for the fitted circuit parameters.

\section{Analytical Matrix Theory of Modulated Bus \label{sec:busmatrixtheory}}
\subsection{Derivation of Matrix Equations}

Here we illustrate in more detail the analytical framework used in the main text to predict the behavior of the EM fields in the tunable Bus resonator with two AC flux driven SQUIDs at the boundaries. We begin with Eqn.~\ref{eqn:eombndry} and fill in the calculations and transformations that yield the matrix equation, Eqn.~\ref{eq:matrixformBus}. Since we are interested in the frequency domain response of the resonance mode field profile, we Fourier transform Eqn.~\ref{eqn:eombndry} to arrive at
\begin{equation}
    \lim_{x \to \pm\frac{\ell}{2}}\left[-\frac{2\omega^2}{\omega_\text{J}^2} \Tilde{\phi} + 2\int dt e^{i\omega t}\cos{f_\pm(t)}\sin{\phi} \pm \eta \ell \frac{\partial\Tilde{\phi}}{\partial x}\right] = 0
    \label{eq:freqboundaryeom}
\end{equation}
with the definition $\Tilde{\phi} (x,\omega) = \int dt e^{i\omega t} \phi(x,t)$ for the Fourier transform.

By linearizing $\sin{\phi} \simeq \phi$ and assuming that the boundary flux modulation takes the form $f_{\pm}(t) = F + \delta f_{\pm} \sin{(\omega_\text{f} t+\psi_{\pm})}$, we can further show that 
\begin{multline}\label{eq:transformMT}
    \int dt e^{i\omega t}\cos{f_{\pm}(t)}\sin{\phi(x,t)} \approx \\
    \sum_{n} \frac{1}{2} e^{in\psi_{\pm}} J_n(\delta f_{\pm}) \left[e^{iF}+ (-1)^{n}e^{-iF}\right]\Tilde{\phi}(x,\omega+n\omega_\text{f})
\end{multline}
where $J_n(\delta f_{\pm})$ is the $n$th order Bessel function of the first kind. Eqns.~\ref{eq:freqboundaryeom}$-$\ref{eq:transformMT} indicate that the parametric drive mixes frequency components of the EM field that are separated by integer multiples of the drive frequency $\omega_{\text{f}}$ in the CPW Bus. Applying a spatial and temporal Fourier transform to Eqn.~\ref{eqn:eombulk} yields the usual linear dispersion relation $\omega = v|k|$. Thus we can decompose $\Tilde{\phi}(x,\omega)$ into left- and right-moving travelling waves,  $\Tilde{\phi}(x,\omega) = \Tilde{\phi}_-(\omega)e^{-ikx}+\Tilde{\phi}_+(\omega)e^{ikx}$.

Inserting $\omega=\omega_{\text{r}}+m \omega_{\text{f}}$ for $m=0, \pm 1, \dots$ into Eqn.~\ref{eq:freqboundaryeom} $-$ \ref{eq:transformMT} we can express the resulting system of equations compactly in matrix form in Fourier space
\begin{equation}
    \label{eq:freqmatrixform}
    \mathbf{M}\mathbf{\Phi}=
    \begin{pmatrix}
    \mathbf{M}_{+,+} & \mathbf{M}_{-,+}\\
    \mathbf{M}_{+,-} & \mathbf{M}_{-,-}
    \end{pmatrix}  
    \begin{pmatrix}
    \mathbf{\Phi}_+\\
    \mathbf{\Phi}_-
    \end{pmatrix}=\mathbf{0}
\end{equation}
where we define the vectors of side-band amplitudes, $\mathbf{\Phi}_\pm$, as
\begin{equation}
    \mathbf{\Phi}_\pm =
    \begin{pmatrix}
    \vdots\\
    \Tilde{\phi}_\pm(\omega_\text{r} - 2\omega_\text{f})\\
    \Tilde{\phi}_\pm(\omega_\text{r} - \omega_\text{f})\\
    \Tilde{\phi}_\pm(\omega_\text{r})\\
    \Tilde{\phi}_\pm(\omega_\text{r} + \omega_\text{f})\\
    \Tilde{\phi}_\pm(\omega_\text{r} + 2\omega_\text{f})\\
    \vdots
    \end{pmatrix}
\end{equation}
where $\omega_{\text{r}}$ is the tunable Bus resonance frequency under parametric driving, to be determined from Eqn.~\ref{eq:freqmatrixform} (see below). Since only the relative phase difference between the boundary flux drives, $\psi_{0}=\psi_{+}-\psi_{-}$, is physical, we set the phases so $\psi_{\pm}=\pm \psi_{0}/2$. To simply the following expressions, we define $k_n = \frac{\omega_n}{v}$. Then, the components of the $\mathbf{M}$-matrices can be written in the compact form
\begin{multline}\label{eq:freqmatrixelement}
     [\mathbf{M}_{s=\pm,z=\pm}]_{mp}=\big[(-\alpha\omega_p^2+sz\eta d k_p i)\delta_{mp}+\\
     (e^{iF}+(-1)^{m-p}e^{-iF})J_{p-m}(\delta f)e^{z i\frac{\psi_0}{2}(p-m)}\big]e^{sz\frac{\omega_p d}{2v}i}.
\end{multline}
\begin{figure}
    \centering
    \includegraphics{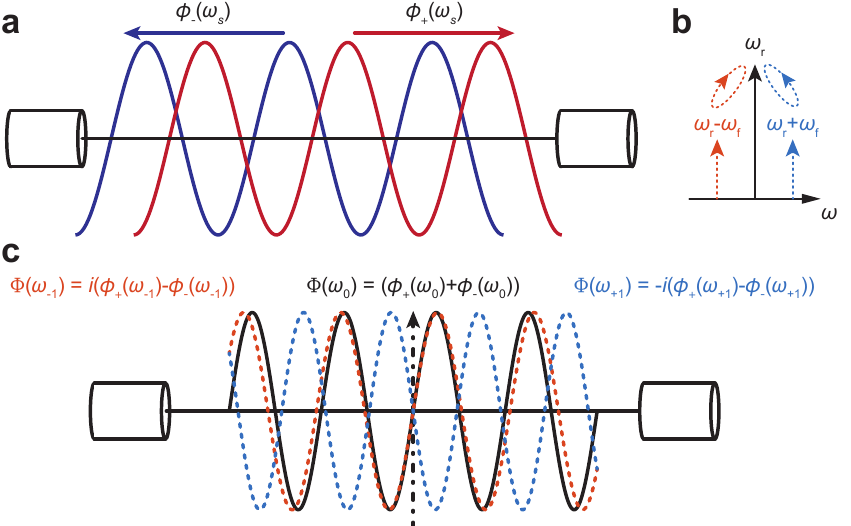}
    \caption{\textbf{a}, Decomposition of sideband-$\omega_{s}$ into left (blue) and right (red) propagating wave frequency components. \textbf{b}, Bus resonant mode $\omega_{\text{0}}$ is coupled to its nearest sidebands separated from it by the boundary flux modulation frequency $\omega_{\text{f}}$. \textbf{c}, Sidebands ($\omega_{\pm}$) imaginary waveforms (dashed blue and orange) associated with the real anti-symmetric 8th order resonant mode (solid black) at $\omega_{0}$. The sidebands generated as the SQUID terminated Bus is under boundary flux modulation are anti-symmetric with respect to the center (vertical dot-dashed black line) of the Bus resonator.}
    \label{fig:fdwaveformdecomp}
\end{figure}

\subsection{Solution Method and Symmetry Transformations \label{ssec:symmetry}}

The parametric drive modified resonance frequency of the Bus, $\omega_\text{r}$, is calculated by solving for $\omega_\text{r}$ such that $\mathrm{det}(\mathbf{M}(\omega_\text{r})) = 0$. The corresponding sideband frequency domain amplitudes can be extracted relative to the resonance mode amplitude by solving for the linear basis of $\mathrm{Null}(\mathbf{M}(\omega_\text{r}))$. Equivalently, the vector of sideband amplitudes is the eigenvector of $\mathbf{M}$ with eigenvalue 0 when $\omega=\omega_{\text{r}}$. It should be pointed out that the frequency component discretization $\Tilde{\phi}_\pm(\omega_n)$ has $n\in \mathbb{Z}$ from $-\infty$ to $+\infty$ with spectral step size $\omega_\text{f}$ corresponding to the Bus SQUID flux modulation frequency. This is clear from Eqn.~\ref{eq:transformMT} which shows that frequency components of the Bus EM field only interact with each other if they are separated by integer multiples of $\omega_\text{f}$. Each resonant mode of the Bus generates a set of non-zero sidebands. Unless two resonant modes are separated by integer multiples of the drive frequency, different sets of sidebands associated with different resonant modes do not interact or mix with each other.

The matrix $\mathbf{M}$ is in principle infinite in size but we could truncate it by considering only a few strong sidebands around the Bus resonance $\omega_\text{r}$, since $|J_n(x)|$ quickly goes to 0 with increasing $|n|$. We truncate the system and consider frequency components with $\omega = \omega_n$ where $n = -N, \hdots, N$. From here, it is straightforward to solve for the modified Bus resonance frequency, $\omega_{\text{r}}$, and corresponding mode profile and the relative sideband amplitudes numerically for the $2(2N+1)$ dimensional matrix equation, Eqn.~\ref{eq:freqmatrixform}. 

This purely numerical treatment can achieve arbitrary accuracy by going to larger $N$ with manageable computational complexity, $O(N^{2})$, but it fails to provide clear physical intuition about the system. To gain further theoretical insights into the Bus dynamics under parametric driving we take advantage of the symmetries of the physical system to simplify Eqn.~\ref{eq:freqmatrixform}. We have already transformed the system into a more symmetric form by translating the spatial coordinate system origin to the center of the Bus and equally splitting the relative flux drive phase, $\psi_0$, between the drives at the two boundaries. The implicit transformation which led to the matrix elements representation in Eqn.~\ref{eq:freqmatrixelement} further leads to clear unitary relations between the sub-matrices, $\mathbf{M}_{\pm,\pm}$:
\begin{equation}\label{eq:freqmatrixSymm}
    \begin{aligned}
    &\mathbf{M}_{-,+} = \mathbf{\Lambda}(\{e^{i(\pi-\psi_0)n}\})\mathbf{M}^*_{+,+}\mathbf{\Lambda}^\dag (\{e^{i(\pi-\psi_0)n}\})\\
    &\mathbf{M}_{+,-} = \mathbf{\Lambda}(\{e^{i\pi n}\})\mathbf{M}^*_{+,+}\mathbf{\Lambda}^\dag (\{e^{i\pi n}\})\\
    &\mathbf{M}_{-,-} = \mathbf{\Lambda}(\{e^{i\psi_0 n}\})\mathbf{M}_{+,+}\mathbf{\Lambda}^\dag (\{e^{i\psi_0 n}\})
    \end{aligned}
\end{equation}
where $\bm{\Lambda}(\{a_{n}\})$ is the $2N+1$ dimensional diagonal matrix with $n$th diagonal element $a_{n}$ for $n=-N,\dots,N$. These relationships highlight the correspondence between the left travelling wave components incident on the right boundary, $\phi_{-}(\omega,\ell/2)$, and the right travelling wave components incident on the left boundary, $\phi_{+}(\omega,-\ell/2)$, as well as the correspondence between the left travelling wave components at the left boundary, $\phi_{-}(\omega,-\ell/2)$, and the right travelling wave components at the right boundary, $\phi_{+}(\omega,\ell/2)$.

We now specialize to the case of $\psi_{0}=0$, with modulation in phase at the two boundaries. We transform the Fourier basis via $\mathbf{\Phi}' = \mathbf{T}^{-1}\mathbf{\Phi}$:
\begin{equation}\label{eq:freqmatrixTransform}
    \mathbf{T} = \begin{pmatrix}
    \mathbf{\Lambda}(\{e^{i\pi n/2}\}) & 0\\
     0 & \mathbf{\Lambda}(\{e^{i\pi n/2}\})
    \end{pmatrix}.
\end{equation}
The Fourier basis transformation simplifies $\mathbf{M}$ to
\begin{equation}
    \label{eq:symmetricM}
    \mathbf{M}' = \mathbf{T}^{-1}\mathbf{M}\mathbf{T}=
    \begin{pmatrix}
        \Sigma & \Sigma^{*} \\
        \Sigma^{*} & \Sigma
    \end{pmatrix}
\end{equation}
where 
\begin{equation}
    \Sigma=\mathbf{\Lambda}(\{e^{-i\pi n/2}\})\mathbf{M}_{+,+}\mathbf{\Lambda}(\{e^{i\pi n/2}\}).
\end{equation}
This transformation rotates the complex sideband amplitudes according to their order so that they all point along the real axis in the complex plane.

$\mathbf{M}'$ can be block diagonalized using the invertible linear map $\mathbf{U}$ such that $\tilde{\mathbf{\Phi}} = \mathbf{U}\mathbf{\Phi}'$ with
\begin{equation}
    \mathbf{U}=
    \begin{pmatrix}
        \mathbf{I} & \mathbf{I}\\
        \mathbf{I} & -\mathbf{I}
    \end{pmatrix}/\sqrt{2}
\end{equation}
and
\begin{equation}\label{eq:finalsymmetricform}
    \tilde{\mathbf{M}} = \mathbf{U}\mathbf{M}'\mathbf{U}^{-1}=
    \begin{pmatrix}
        \mathbf{\Sigma} + \mathbf{\Sigma}^* & 0\\
        0 & \mathbf{\Sigma} - \mathbf{\Sigma}^*
    \end{pmatrix}.
\end{equation}
This transformation explicitly decouples the subspaces of Bus modes that are spatially even and odd, highlighting the spatial inversion symmetry of the physical system. Note that all the transformations used here are invertible transformations and they preserve $\mathrm{Null}(\mathbf{M})$. 

As noted earlier, each resonant mode of the CPW Bus generates a set of sidebands separated by $\omega_\text{f}$ under the parametric drive. As long as there is no $n' \in \mathbb{Z}$ such that $n'\omega_\text{f}=|\omega_\text{r} - \omega_\text{r}'|$ then the unique modes of the Bus are decoupled and can be analyzed independently. As a result, for each resonance $\omega_\text{r}$ (determined by $\mathrm{det}(\mathbf{M}(\omega_\text{r}))=0$), the necessary requirement for the validity of the system is $\mathrm{dim}(\mathrm{Null}(\Tilde{\mathbf{M}}(\omega_{\text{r}})) = \mathrm{dim}(\mathrm{Null}(\mathbf{M}(\omega_{\text{r}})) = 1$. Given Eqn.~\ref{eq:finalsymmetricform}, this requirement indicates that either $\mathrm{det}(\mathbf{\Sigma} + \mathbf{\Sigma}^*) = 0, \mathrm{det}(\mathbf{\Sigma} - \mathbf{\Sigma}^*) \neq 0$ or $\mathrm{det}(\mathbf{\Sigma} + \mathbf{\Sigma}^*) \neq 0, \mathrm{det}(\mathbf{\Sigma} - \mathbf{\Sigma}^*) = 0$ for some $\omega_{\text{r}}$. The former indicates that $\Tilde{\Phi}_- = \Tilde{\Phi}_+$ and the latter indicates that $\Tilde{\Phi}_- = -\Tilde{\Phi}_+$. These two scenarios correspond to spatially symmetric wave solutions and spatially anti-symmetric wave solutions, respectively.

Spatial symmetry of the modes transitions as a function of boundary bias when $Z(F) \approx Z_{0}$. For the measurements in Fig.~\ref{fig:parametricqubitbus} and Fig.~\ref{fig:q14exchange} we fix the operating point at $F=\pi/4$ and the modes have the same spatial symmetry as they do at $F=0$ since $Z(\pi/4) < Z_{0}$. Thus, since we are interested in the behavior of the 8th order mode which dominantly mediates the coupling, we solve $(\Sigma - \Sigma^{*})\tilde{\mathbf{\Phi}}_{-}=0$ from this point on.

So far we have simplified the linear system dimension two-fold and identified the solutions' symmetry properties by exploring the algebraic properties of the system. We could further reduce the number of degrees of freedom by observing that $(\mathbf{\Sigma} + \mathbf{\Sigma}^*)$ and $(\mathbf{\Sigma} - \mathbf{\Sigma}^*)$ can be made transpose symmetric by transforming the Fourier basis as $\tilde{\mathbf{\Phi}}' = \mathbf{Q}^{-1}\Tilde{\mathbf{\Phi}}$ with
\begin{equation}\label{eq:finaltransposeSym}
    \mathbf{Q} = 
    \begin{pmatrix}
        \mathbf{\Lambda}(\{\frac{1}{\cos{(\frac{\pi n}{2}-\frac{\omega_n d}{2v})}} \}) & 0\\
         0 & \mathbf{\Lambda}(\{\frac{1}{\sin{(\frac{\pi n}{2}-\frac{\omega_n d}{2v})}}\})
    \end{pmatrix}.
\end{equation}
This scales the side-band amplitudes to emphasize the symmetry between the $\tilde{\phi}(\omega_{\text{r}}+m\omega_{\text{f}})$ and $\tilde{\phi}(\omega_{\text{r}}-m\omega_{\text{f}})$ sidebands.

\subsection{Derivation of Coupling from Sideband Amplitudes}

In practice, we use the transformations leading up to Eqn.~\ref{eq:finalsymmetricform} in our numerical solutions for $\mathbf{\tilde{\Phi}}$ (shown in the inset of Fig.~\ref{fig:parametricqubitbus}\textbf{a} as a function of boundary flux modulation amplitude, $\delta f$, and frequency, $\omega_\text{f}$). We have also numerically verified the validity of these transformations by comparing the results to a direct solution of Eqn.~\ref{eq:freqmatrixform} in certain test cases. In order to arrive at Eqn.~\ref{eqn:gpar} we use the Fourier transformed sideband weights obtained from solving Eqn.~\ref{eq:matrixformBus} to compute the spatiotemporal phase field mode profile, $\phi(x,t)$. Here we assume that, since the flux drive amplitude is small, $\omega_{\text{r}}$ is not substantially changed from its value in the absence of the drive. Thus, we take $\omega_{\text{r}}=\pi N\frac{v}{\ell}$ ($N=1,2,\dots$) for modes of the Bus resonator coupled to $N$ qubits. Under this set of assumptions, the phase field $\tilde{\phi}(x,\omega_{\pm})$ can be calculated at the position of qubit $m$, $x=x_{m}=(m+1/2)\ell/N$, for $m=-N/2,\dots,N/2 -1$:
\begin{align}
    \tilde{\phi}(x_{m},\omega_{\pm})&=\tilde{\phi}_{-}(\omega_{\pm})e^{-i\omega_{\pm}x_{m}/v}+\tilde{\phi}_{+}(\omega_{\pm})e^{i\omega_{\pm}x_{m}/v} \notag \\
    &=i(-1)^{m}\left[\tilde{\phi}_{+}e^{\pm i\omega_{f}x_{m}/v}-\tilde{\phi}_{-}e^{\mp i\omega_{f}x_{m}/v}\right] \notag \\
    & \approx 2i(-1)^{m}\cos\left(\frac{\omega_{\text{f}}x_{m}}{v}\right)\tilde{\phi}(\omega_{\pm}).
    \label{eqn:sidebandprofile}
\end{align}
where we consider the experimentally relevant case of odd sidebands and central band:
\begin{equation}
    \tilde{\phi}_{+}(\omega_{p})=-\tilde{\phi}_{-}(\omega_{p}),\ \ \ p=0,\pm 1.
\end{equation}
In the limit of low modulation frequency, $\omega_{\text{f}} \ll \omega_{\text{FSR}}$ we have $\omega_{\text{f}}x_{m}/v \ll 1$ and the first order sidebands achieve their maxima at the the qubit positions, as desired. As $\omega_{f}$ increases the sideband amplitude at the qubit positions is reduced by a small amount, which contributes to the parametric coupling decreasing with modulation frequency (see Fig.~\ref{fig:parametricqubitbus}\textbf{a}). It should be noted that this contribution is small and the decreasing parametric coupling with modulation frequency is dominated by the sideband Fourier amplitudes themselves decreasing with modulation frequency. 
We can generalize the expression for the qubit-Bus coupling from Eqn.~\ref{eqn:gpar} to write the time dependent qubit-Bus coupling as $g_{m}(t)=C_{m,\text{r}}V_{0,m}V_{\text{mod},\text{r}}(x_{m},t)$ in terms of the Bus voltage profile in the presence of boundary modulation, $V_{\text{mod},\text{r}}(x_{m},t)$. The voltage and phase fields are related by a time derivative, so using Eqn.~\ref{eqn:sidebandprofile} we calculate
\begin{multline}
    \frac{V_{\text{mod},\text{r}}(x_{m},t)}{V_{0,\text{r}}(x_{m})}=\frac{\omega_{\text{r}}}{\omega_{\text{in},\text{r}}}\frac{(-1)^{m}}{\tilde{\phi}_{\text{in}}(\omega_{\text{in},\text{r}})} \notag \\
    \times\left\{\tilde{\phi}(\omega_{\text{r}})+i\cos\left(\frac{\omega_{\text{f}}x_{m}}{v}\right)\sin(\omega_{\text{f}}t)\left[\tilde{\phi}(\omega_{+})-\tilde{\phi}(\omega_{-})\right]\right\}
\end{multline}
where $\omega_{\text{in},\text{r}}$ and $\tilde{\phi}_{\text{in}}$ are the frequency and central band amplitude, respectively, in the absence of boundary modulation of the Bus.

Thus we have shown that the central band contributes a static term to the coupling while the sidebands contribute a time periodic term oscillating at the modulation frequency:
\begin{equation}
    \label{eqn:timeperiodiccoupling}
    g_{m}(t)=g_{0m}\left[\bar{g}_{0m} + 2\bar{g}_{m}\sin(\omega_{\text{f}}t)\right]
\end{equation}
where
\begin{equation}
    \bar{g}_{0m}=\frac{\omega_{\text{r}}}{\omega_{\text{in},\text{r}}}\frac{\tilde{\phi}(\omega_{\text{r}})}{\tilde{\phi}_{\text{in}}(\omega_{\text{in},\text{r}})}
\end{equation}
and
\begin{equation}
    \bar{g}_{m}=\frac{i}{2}\left[\frac{\tilde{\phi}(\omega_{+})-\tilde{\phi}(\omega_{-})}{\tilde{\phi}_{\text{in}}(\omega_{\text{in},\text{r}})}\right]\cos\left(\frac{\omega_{\text{f}}x_{m}}{v}\right).
\end{equation}
In order to reproduce Eqn.~\ref{eqn:gpar} we note that (from above)
\begin{align}
    \tilde{\mathbf{\Phi}}&=\mathbf{U}\mathbf{T}^{-1}\mathbf{\Phi} \notag \\
    &=\frac{1}{\sqrt{2}}
    \begin{pmatrix}
        \left\{e^{-i\pi m/2}\left[\tilde{\phi}_{+}(\omega_{\text{r}}+m\omega_{f})+\tilde{\phi}_{-}(\omega_{\text{r}}+m\omega_{f})\right]\right\} \\[5pt]
        \left\{e^{-i\pi m/2}\left[\tilde{\phi}_{+}(\omega_{\text{r}}+m\omega_{f})-\tilde{\phi}_{-}(\omega_{\text{r}}+m\omega_{f})\right]\right\}
    \end{pmatrix}
    \label{eqn:defphitilde}
\end{align}
and we solve the lower right block of $\mathbf{\Tilde{M}}\mathbf{\tilde{\Phi}}=0$ for spatially odd modes.

\subsection{Energy Normalization of Sideband Amplitudes}

In order to achieve the quantitative agreement between theoretical and experimental coupling rates demonstrated in Fig.~\ref{fig:parametricqubitbus}\textbf{a} and Fig.~\ref{fig:q14exchange}\textbf{b} we need to properly normalize the sideband amplitudes computed in the previous sections. The sideband amplitude vector $\mathbf{\Phi}$ should be normalized such that the energy stored in the zero point EM field is preserved when the boundary flux modulation is turned on. Since the total zero point energy is $(1/2)\hbar\omega_{0,\text{r}}$ and the energy is stored equally in the electric and magnetic fields, the component stored in the electric field of the unmodulated mode is
\begin{align}
    \frac{1}{4}\hbar\omega_{0,\text{r}}&=\frac{C_{0}}{2}\int_{-\ell/2}^{\ell/2}|V_{0,\text{r}}(x)|^{2}dx \notag \\
    &=\frac{C_{0}\ell}{4}\omega_{0,\text{r}}^{2}|\tilde{\phi}_{\text{in},0}|^{2}
    \label{eqn:normnomod}
\end{align}
where as a shorthand we write $\phi_{m}\equiv \phi_{+}(\omega_{m})=-\phi_{-}(\omega_{m})$ for the sideband amplitudes and again we take the experimentally relevant case of spatially odd modes. In the presence of modulation we must include the contribution to the energy due to the sidebands:
\begin{align}
    \frac{1}{4}\hbar\omega_{0,\text{r}}&=\frac{C_{0}}{2}\int_{-\ell/2}^{\ell/2}|V_{\text{mod},\text{r}}(x)|^{2}dx \notag \\
    &=\frac{C_{0}\ell}{4}\sum_{m}\omega_{m}^{2}|\tilde{\phi}_{m}|^{2}\left[1+(-1)^{m}\sin\left(\frac{\omega_{m}\ell}{v}\right)\right].
    \label{eqn:normyesmod}
\end{align}
Equating Eqn.~\ref{eqn:normnomod} and Eqn.~\ref{eqn:normyesmod} allows for a straightforward substitution of the numerical results into Eqn.~\ref{eqn:gpar} to compute the parametric coupling rates.

\section{Multi-Tone Frequency Domain Experiments \label{sec:freqdomain}}
The state of the Bus resonator cannot be probed directly so its spectrum in Fig.~\ref{fig:spidernetcircuit}\textbf{b} is measured by using a vector network analyzer (VNA) to monitor a weak probe tone at the frequency of one of the qubit readout resonators while sweeping a second tone sent to the Bus drive line. The chosen readout resonator's qubit acts as a spectrometer coupled to the Bus resonator. When the Bus is driven on resonance with one of its modes, the response of the spectrometer qubit is observed through its readout resonator. Resulting parameters from fitting the measured multi-mode spectrum of the bus are found in Table~\ref{tab:DCfitpars} as discussed in the main text.

In order to extract the frequency domain avoided crossings between a qubit and the Bus (like Fig.~\ref{fig:parametricqubitbus}\textbf{b}) we employ a similar multi-tone experiment. Again, a weak probe tone at the readout resonator of Q$_i$ monitors its response. A second tone to the qubit charge drive line is swept across the resonance frequency of the qubit, $\omega_{\text{q},i}$. Finally, phase correlated flux modulation tones are sent to the boundary SQUIDs with amplitude $\delta f$, frequency $\omega_{\text{f}}$ and relative phase $\psi_{0}=0$, which maximizes the sideband amplitudes and thus the coupling at the chosen static bias point. See the level diagram in Fig.~\ref{fig:parametricqubitbus}\textbf{f}.
\begin{table}
    \centering
    \begin{tabular}{|c|c|c|}
        \hline
        Parameter & Predicted & Fit (rel. to predicted) \\
        \hline
        \hline
        $E_{\mathrm{C}}/h$ & 9.68 GHz & 0.202 \\
        $E_{\mathrm{J}}/h$ & 397 GHz & 1.11 \\
        $E_{\mathrm{L}_0}/h$ & 5.11 GHz & 0.757 \\
        $v$ & 0.49$c$ & 1.26 \\
        $d$ & 0 & 0.0663 \\
        \hline
    \end{tabular}
    \caption{Free parameters in the theoretical model of the frequency spectrum of the tunable Bus. The fit value is given as the ratio of the fit parameter to its predicted value, except in the case of the SQUID asymmetry, $d$, where the SQUIDs are designed to be symmetric but junction to junction critical current variations during fabrication introduce a small asymmetry. The speed of light in the bulk CPW, $v$ is measured in terms of the free space speed of light, $c$. $E_{\mathrm{C}}$ fit from experiment is noticeably smaller than its predicted value. Since this device utilizes flip-chip integration there is additional ground metal immediately above the junction on the facing chip which may introduce the additional capacitance responsible for lowering $E_{\mathrm{C}}$ \cite{Kosen_2022}.}
    \label{tab:DCfitpars}
\end{table}

\section{Detailed tSWT for Qubit-Qubit Dynamics \label{sec:tswt}}
Here we elaborate on the calculation that yields Eqn.~\ref{eq:Hinteff} starting from an initial Hamiltonian composed of terms
\begin{subequations}
    \label{eq:Hfull}
    \begin{align}
        H_{0}/\hbar&=\omega_{\text{r}}a^{\dagger}a+\sum_{i}\frac{1}{2}\omega_{i}\sigma_{i,z}, \label{eq:H0} \\
        H_{1}/\hbar&=-\sum_{i=1,2}g_{i}(t)\left(\sigma_{i,-}-\sigma_{i,+}\right)\left(a-a^{\dagger}\right). \label{eq:H1}
    \end{align}
\end{subequations}
We employ a tSWT where the Hamiltonian $H=H_{0}+H_{1}$ is transformed according to
\begin{equation}
    H'=e^{S(t)}He^{-S(t)}+i\frac{\partial e^{S(t)}}{\partial t}e^{-S(t)},
\end{equation}
where the time dependence of $S(t)$ is emphasized. The goal is to find a generator for the transformation, $S(t)$, such that the resulting Hamiltonian $H'$ is block-diagonal with respect to the Bus resonator state. The terms in $H'$ can be expanded as
\begin{equation}
    e^{S}He^{-S}=H+[S,H]+\frac{1}{2!}\left[S,[S,H]\right]+\dots
\end{equation}
and
\begin{equation}
    i\frac{\partial e^{S}}{\partial t}e^{-S}=i\left(\dot{S}+\frac{1}{2!}[S,\dot{S}]+\frac{1}{3!}[S,[S,\dot{S}]]+\dots\right)
\end{equation}
using the Baker-Campbell-Hausdorff formula. Terms are then grouped by order,
\begin{align}
    H_{\text{eff}}^{(0)}&=H_{0} \\
    H_{\text{eff}}^{(1)}&=i\dot{S}_{1}+[S_{1},H_{0}]+H_{1} \\
    H_{\text{eff}}^{(2)}&=i\dot{S}_{2}+\frac{i}{2}[S_{1},\dot{S}_{1}]+\frac{1}{2}[S_{1},[S_{1},H_{0}]]+[S_{1},H_{1}].
\end{align}
where $S=\sum_{n}S_{n}$ is decomposed by order. At lowest order we consider just $S_{1}$. Borrowing inspiration from the well known application of the SWT to the fixed coupling between a qubit and resonator \cite{Blais2020Review}, we guess the form of $S_{1}$ is
\begin{equation}
    S_{1}=\sum_{i=1,2}\left(c_{i}a^{\dagger}\sigma_{i,-}-c_{i}^{*}a\sigma_{i,+}+d_{i}a^{\dagger}\sigma_{i,+}-d_{i}^{*}a\sigma_{i,-}\right)
\end{equation}
where the coefficients $c_{i}$ and $d_{i}$ will be time dependent in general.

These coefficients are determined by enforcing the condition $H_{\text{eff}}^{(1)}=0$ which produces the set of EOMs for the coefficients,
\begin{align}
    \dot{c}_{i}&=-i\Delta_{i}c_{i}+ig_{i} \\
    \dot{d}_{i}&=-i\Sigma_{i}d_{i}-ig_{i}.
\end{align}
These equations are readily formally integrated
\begin{align}
    c_{i}(t)&=i\int_{0}^{t}dt'g_{i}(t')e^{i\Delta_{i}(t'-t)} \\
    d_{i}(t)&=-i\int_{0}^{t}dt'g_{i}(t')e^{i\Sigma_{i}(t'-t)}.
\end{align}
using $g_{i}(t)$ from Eqn.~\ref{eqn:timeperiodiccoupling} to obtain solutions
\begin{align}
    \label{eq:ci}
    c_{i}(t)&=\frac{g_{0i}}{\Delta_{i}}+\frac{g_{0i}}{(\Delta_{i}-\omega)(\Delta_{i}+\omega)}\big[\Delta_{i}\cos(\omega t+\phi) \notag \\
    &\ \ \ \ \ -i\omega \sin(\omega t+\phi)-e^{-i\Delta_{i}t}\left(\Delta_{i}\cos\phi -i\omega\sin\phi\right)\big] \\
    \label{eq:di}
    d_{i}(t)&=-\frac{g_{0i}}{\Sigma_{i}}+\frac{g_{0i}}{(\Sigma_{i}-\omega)(\Sigma_{i}+\omega)}\big[-\Sigma_{i}\cos(\omega t+\phi) \notag \\
    &\ \ \ \ \ +i\omega \sin(\omega t+\phi)+e^{-i\Sigma_{i}t}\left(\Sigma_{i}\cos\phi-i\omega\sin\phi\right)\big].
\end{align}
Given the choice of coefficients $c_{i}$ and $d_{i}$ we have the relation $H_{1}=-i\dot{S}_{1}-[S_{1},H_{0}]$ and the second order term in the resulting effective Hamiltonian simplifies to
\begin{equation}
    H_{\text{eff}}^{(2)}=-\frac{1}{2}[S_{1},H_{1}].
\end{equation}
This is straightforward to compute, and in terms of the $c_{i}$ and $d_{i}$ the transformed Hamiltonian becomes, through second order,
\small
\begin{align}
    \label{eqn:fulltransformH}
    H'&=a^{\dagger}a\Big[\omega_{\text{r}}+\underbrace{\frac{1}{2}\sum_{i=1,2}g_{i}(t)\sigma_{i,z}(c_{i}+c_{i}^{*}+d_{i}+d_{i}^{*})}_{\text{AC Stark Shift}}\Big] \notag \\
    &\ \ \ +\sum_{i=1,2}\frac{1}{2}\sigma_{i,z}\Big[\omega_{i}+\underbrace{\frac{3}{2}g_{i}(t)(c_{i}+d_{i}+c_{i}^{*}+d_{i}^{*})}_{\text{Lamb Shift}}\Big] \notag \\
    &\ \ \ -\underbrace{\frac{1}{4}\sum_{i=1,2}g_{i}(t)(-c_{i}-c_{i}^{*}+d_{i}+d_{i}^{*})}_{\text{overall energy shift}} \notag \\\
    &\ \ \ -\underbrace{\frac{1}{2}\left(a^{\dagger}\right)^{2}\sum_{i=1,2}\left[g_{i}(t)\sigma_{i,z}(c_{i}+d_{i})+\text{h.c.}\right]}_{\text{state dependent squeezing}} \notag \\
    &\ \ \ -\underbrace{\sigma_{1,-}\sigma_{2,+}\frac{1}{2}\left[g_{1}(t)(d_{2}-c_{2}^{*})+g_{2}(t)(-c_{1}+d_{1}^{*})\right]+\text{h.c.}}_{\text{photon exchange interaction}} \notag \\
    &\ \ \ -\underbrace{\sigma_{1,+}\sigma_{2,+}\frac{1}{2}\left[g_{1}(t)(c_{2}^{*}-d_{2})+g_{2}(t)(c_{1}^{*}-d_{1})\right]+\text{h.c.}}_{\text{two photon creation/annihilation interaction}}
\end{align}
\normalsize
Next we insert the expressions for $c_{i}(t)$ and $d_{i}(t)$ from Eqns.~\ref{eq:ci} $-$ \ref{eq:di} with $\omega=\tilde{\Delta}_{12}$. Lastly, Eqn.~\ref{eq:Hinteff} is recovered by moving to the interaction frame and making the rotating wave approximation, keeping only the static terms in the interacting frame.

\section{Parametric Two-Qubit Gate Calibration and Benchmarking\label{sec:2qgatecal}}
\begin{figure}
    \centering
    \includegraphics{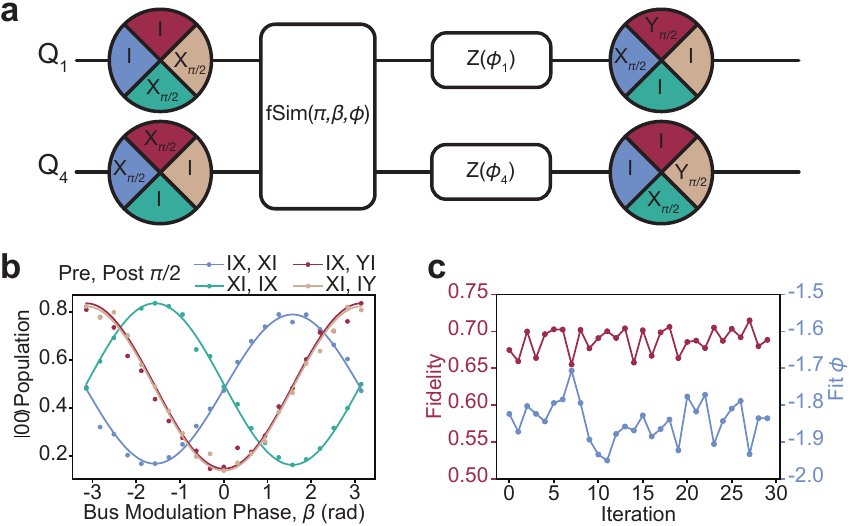}
    \caption{\textbf{a}, Circuit for calibration of post modulation single qubit Z rotations. The first pair of $\pi/2$ gates applied to both qubits initialize them along the $(Z,Y)$ or $(Y,Z)$ axes. For $(Z,Y)$ ($(Y,Z)$) a set of final tomography pulses allows a measurement along the $(Y,Z)$ or $(X,Z)$ ($(Z,Y)$ or $(Z,X)$) axes. In between we apply the resonant parametric modulation tone sweeping the modulation phase $\beta$. \textbf{b}, The single qubit Z rotation angles $\phi_{1}$ and $\phi_{4}$ are chosen to shift the data in phase with the solid curves for an ideal $\text{fSim}(\pi,0,\phi)$ gate. \textbf{c}, Repeated two-qubit $\text{fSim}$ process fidelity measurements. QPT is repeated 31 times over 12 hours to investigate the fluctuations of the process fidelity (red, left axis) and fitted conditional phase (blue, right axis). The upper plot limit represents the coherence limited process fidelity, $\mathcal{F}=0.753$.}
    \label{fig:gatecalib}
\end{figure}

The demonstrated parametric pairwise coherent photon exchange naturally implements a Fermionic simulation (fSim) type interaction (see Eqn.~\ref{eqn:fulltransformH}) \cite{Foxen2020fSim},
\begin{equation}
    \label{eqn:fsimunitary}
    \text{fSim}(\theta,\beta,\phi)=
    \begin{pmatrix}
        1 & 0 & 0 & 0 \\
        0 & \cos\left(\frac{\theta}{2}\right) & i\sin\left(\frac{\theta}{2}\right)e^{i\beta} & 0 \\
        0 & i\sin\left(\frac{\theta}{2}\right)e^{-i\beta} & \cos\left(\frac{\theta}{2}\right) & 0 \\
        0 & 0 & 0 & e^{i\phi}
    \end{pmatrix}.
\end{equation}
The iSWAP angle $\theta$ is the rotation angle in the single excitation manifold of the two interacting qubits ($\{\ket{01},\ket{10}\}$), $\beta$ defines the transverse rotation axis within the single excitation manifold and $\phi$ gives the conditional phase due to residual static or dynamic $ZZ$ interactions \cite{Abrams2020}. We fix the Bus modulation amplitude and obtain $\theta=\pi$ by appropriately adjusting the Bus modulation time and frequency. From a time-domain photon exchange experiment like Fig.~\ref{fig:q14exchange}\textbf{a}, the optimal gate time and frequency for achieving $\theta=\pi$ for Q$_{1}$ and Q$_{4}$ are identified to be $\tau = 831~\text{ns}$ and $\omega_{\text{f}}/2\pi=83~\text{MHz}$ respectively. Following that, we apply single qubit Z rotations to both qubits to set $\beta = 0$ \cite{Abrams2020} (see Fig.~\ref{fig:gatecalib}\textbf{a}-\textbf{b} for calibration protocol). The remaining uncalibrated parameter is the conditional phase $\phi$ resulting from a combination of residual static and dynamic $ZZ$ interactions that we do not attempt to suppress or cancel out in this work (see the discussion below). 
We benchmark the resulting fSim gate using quantum process tomography (QPT) \cite{OBrien_QPT}. The QPT is repeated for 30 iterations over a 12-hour period (see Fig.~\ref{fig:gatecalib}\textbf{b}). Over this time interval we observe a stable process fidelity of $\mathcal{F}=68.8 \pm 1.6\%$. The fidelity is determined by choosing the conditional phase $\phi$ which maximizes the measured $\mathcal{F}$ for each iteration. Again we observe a stable optimal conditional phase $\phi=-1.84 \pm 0.055$. Optimizing the fidelity over the other parameters ($\theta$ and $\beta$) in Eqn.~\ref{eqn:fsimunitary} yields $\theta\approx \pi$ and $\beta \approx 0$ with a negligible improvement in the optimized process fidelity so we conclude that the uncalibrated conditional phase represents the dominant source of coherent error with respect to implementing a pure $i\text{SWAP}=\text{fSim}(\pi,0,0)$. We expect that most of this unwanted $ZZ$ can be suppressed by redesigning the device such that it can be operated at the Bus bias point where the bare qubit-qubit coupling $g_{14}\approx0$, thus yielding a static $ZZ \approx 0$. As justification, we measure the $ZZ$ rate as the Bus bias approaches $F=\pi/2$ and compare the results to predictions from a theoretical multi-mode photonics model of the tunable Bus \cite{Sato2012MultiModePhotonics}. The details of the measurement and analysis can be found in SI Note~\ref{sec:mmZZ} where we show that this model agrees well with the data and predicts that the $ZZ$ rate should indeed be negligible when we extrapolate to the designed zero coupling point at $F=\pi/2$.

When the Bus is modulated, coherence times of the qubits are reduced with respect to their base values (see SI Note~\ref{sec:devicepars}) by as much as a factor of 4. These reduced coherence times are used to estimate the coherence limited process fidelity $\mathcal{F}_{\text{coh}}=0.753$ \cite{Dawkins_CohLimit}. Due to the relatively short qubit coherence times and readout linewidth $\kappa$ and dispersive shift $\chi$ that are not optimal we observe substantial readout error (see SI Note~\ref{sec:rcorr}). As a result, the populations shown here are all corrected using the full two qubit readout confusion matrix. This corrects most of the readout errors. However, we observe that there are some significant fluctuations in the elements of the confusion matrix over time. We re-measured the confusion matrix before each of the 16 sequences required for a single QPT iteration and observed fluctuations even at this time scale. Since QPT is a state preparation and measurement (SPAM) dependent benchmarking method we believe that most of the remaining process infidelity can be attributed to measurement error. 

\section{Multi-mode Model for Static \texorpdfstring{$ZZ$}{ZZ} Interactions \label{sec:mmZZ}}
\begin{figure}
    \centering
    \includegraphics{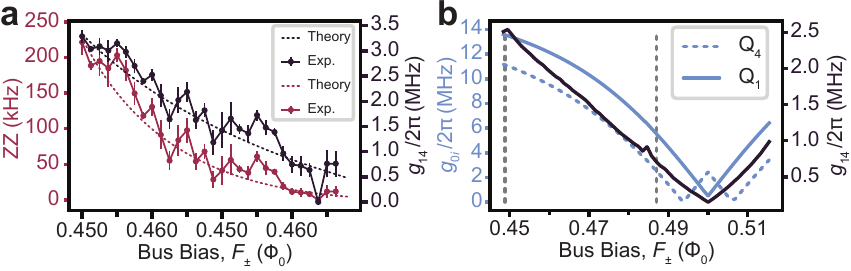}
    \caption{\textbf{a}, Experimental characterization of ZZ rate vs. Bus boundary bias. We use JAZZ to measure the ZZ rate between Q$_{1}$ and Q$_{4}$ (left axis, red) and from this infer the qubit-qubit coupling, $g_{14}$ (right axis, dark purple). Points are data with error bars giving $\pm 1$ standard deviation error on the measurement. The dashed curves are theory based on the calculations in SI Note~\ref{sec:mmZZ}. \textbf{b}, Similarly, the qubit-bus couplings $g_{0i}$ (left axis, light blue) and qubit-qubit coupling $g_{14}$ (right axis, dark purple) can be calculated theoretically in the ideal case of fixed qubit-bus detunings where both quantities go to zero near $F_{\pm}=\pi/2$ as designed. Vertical dashed grey lines correspond to the bounds of the Bus bias range probed experimentally in \textbf{a}.}
    \label{fig:zzextract}
\end{figure}

As discussed in Sec.~\ref{sec:2qgatecal}, unwanted $ZZ$ coupling between Q$_{1}$ and Q$_{4}$ prevents us from implementing a pure $i$SWAP gate. This is a consequence of the fact that flux noise affecting the Bus prevented operation of the device at the intended bias point of $F=\pi/2$ where the bare qubit-Bus and qubit-qubit couplings are designed to be $g_{0i}=g_{ij}=0$. In order to infer that operation of the device at the zero coupling point can yield $ZZ$-free interactions we use Joint-Amplified ZZ (JAZZ), \cite{GARBOWJazz2,TakitaJazz1}, to characterize the $ZZ$ rate between Q$_1$ and Q$_4$ as a function of the Bus flux bias point near $F=\pi/2$. The results are shown in Fig.~\ref{fig:zzextract}\textbf{a} where we plot both the $ZZ$ rate, $\zeta_{14}$, and the inferred $g_{14}$ calculated from
\begin{equation}
    \label{eqn:qubitZZ}
    \zeta_{14}=2g_{14}^{2}\left(\frac{1}{\Delta_{14}-\alpha_{1}}-\frac{1}{\Delta_{14}+\alpha_{4}}\right)
\end{equation}
where $\Delta_{14}=\omega_{1}-\omega_{4}$ is the qubit-qubit detuning and $\alpha_{i}$ is the anharmonicity of qubit $i$. Here we observe the desired trend of decreasing $ZZ$ and thus $g_{14}$ as we move towards $F=\pi/2$. However, these measurements alone are inconclusive since the qubit-qubit coupling rate reduction can be due to both decreasing qubit-Bus coupling and increasing qubit-Bus detuning. A fixed detuning between the qubits and Bus cannot be maintained because further biasing the qubits away from their flux insensitive points to track the Bus frequency reduces their $T_{2e}$ such that we cannot resolve the small $ZZ$ rates near $F=\pi/2$. 

In order to reach a more definitive conclusion, we adapt the multi-mode photonics model developed in \cite{Sato2012MultiModePhotonics} to our system. We will present the essential adaptations of the model to the present system here, but refer to the original work for a more complete presentation of the calculations. The main result is an expression for the qubit-qubit coupling, mediated by a multi-mode resonator, relevant for the current situation where the qubits couple to a shared tunable Bus with a small FSR, $\omega_{\text{FSR}}/2\pi \sim 660~\text{MHz}$. The coupling is
\begin{equation}
    g_{14}=\frac{1}{2}\sqrt{g_{01}g_{04}}\cos\left(\frac{\phi_{1}}{2}\right)\cos\left(\frac{\phi_{4}}{2}\right)\left(\frac{1}{\sin\theta_{1}}+\frac{1}{\sin\theta_{4}}\right)
\end{equation}
where $\phi_{i}$ and $\theta_{i}$ are the round trip propagation phase of a photon at frequency $\omega_{i}$ traveling between qubit $i$ and the Bus boundary and the end to end propagation phase of a photon traveling between the two boundaries of the Bus, respectively. These phases fully characterize the EM mode profile in the Bus and are defined more explicitly as
\begin{equation}
    \phi_{i}=\omega_{i}\frac{2d_{i}}{v}+\varphi(F)
\end{equation}
\begin{equation}
    \theta_{i}=\omega_{i}\frac{\ell}{v}+\varphi(F)
\end{equation}
where $d_{i}$ is the distance between qubit $i$ and the boundary of the Bus, $\ell$ is the full length of the Bus, and $\varphi$ represents the extra phase shift obtained when a photon reflects off the Bus boundary. In general $\varphi$ depends on the impedance boundary condition with $\varphi=\pi$ and $\varphi=0$ for short and open boundary conditions respectively. In the current system, the external Bus flux, $F$ tunes the boundary SQUID impedance and thus tunes $\varphi$ between these extremes in the ideal case. For the SQUID terminated CPW we have
\begin{equation}
    \varphi(F)=\text{arg}\left\{\frac{1+i\frac{\omega_{\text{r}}}{v}L_{\text{eff}}(F)}{1-i\frac{\omega_{\text{r}}}{v}L_{\text{eff}}(F)}\right\}
\end{equation}
where the effective length, $L_{\text{eff}}$ is
\begin{equation}
    L_{\text{eff}}(F)=\frac{L_{\text{SQ}}(F)}{L_{0}}
\end{equation}
the ratio of the SQUID inductance to the bulk CPW inductance per unit length \cite{JohanssonCasimir1}. Using the fitted circuit parameters in Table~\ref{tab:DCfitpars} and the measured qubit frequencies at each $F$ we can compute $\varphi$, $\phi_{i}$, and $\theta_{i}$ which yields both the qubit-qubit coupling, $g_{14}$, (from which the $ZZ$ rate can be inferred using Eqn.~\ref{eqn:qubitZZ}) and the qubit-Bus coupling $g_{0i}$:
\begin{equation}
    g_{0i}(F)=g_{0i}(0)\cos\left(\frac{\phi_{i}}{2}\right)
\end{equation}
where $g_{0i}(0)$ is the qubit-Bus coupling at $F=0$ (where the qubit is located at an antinode of the 8th order Bus mode voltage profile) which we have referred to as simply $g_{0i}$ throughout the main text for simplicity. We first validate the model by using it to predict $g_{14}$ and $\zeta_{14}$ for the scenario we probed experimentally: qubit-Bus detunings changing with changing flux bias $F$. The results are shown in Fig.~\ref{fig:zzextract}\textbf{a} where we obtain very strong agreement with the experimentally measured values over the given range of $F$. 

Having demonstrated the predictive power of the model, we now investigate theoretically the ideal scenario of interest: qubit-Bus detunings fixed with changing flux bias $F$. Here we isolate the effect of $g_{0i}$ decreasing as $F$ is tuned toward $F=\pi/2$. For this analysis we use the experimentally measured Bus frequency as a function of $F$ and fix the qubit-Bus detunings to the measured values for the operating point where we calibrate the two-qubit gate in Sec.~\ref{sec:2qgatecal}. The results are shown in Fig.~\ref{fig:zzextract}\textbf{b}. First, we observe that $g_{0i}$ reaches a zero near $F=\pi/2$ for both qubits. Most importantly, we see that $g_{14}$ (and thus $\zeta_{14}$) approaches zero at $F=\pi/2$ (even with the qubit-Bus detuning fixed), as designed. This provides strong evidence that we are able to design the zero coupling and zero $ZZ$ point accurately. Thus, future devices can be reliably designed such that the zero coupling operation point occurs at a flux bias, $F$, with less flux noise sensitivity, opening up the potential for $ZZ$ free $i$SWAP gates in the proposed architecture.

\section{Readout Correction \label{sec:rcorr}}
\begin{figure}
    \centering
    \includegraphics{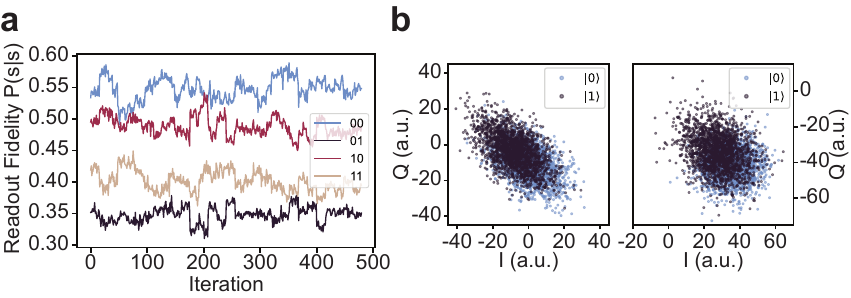}
    \caption{\textbf{a}, The measured confusion matrix components, $C_{ii}$, over time. The legend labels give the state $i$. This shows relatively large random fluctuations that limit our readout fidelity and stability in the particular device used in this work. \textbf{b}, The measured Q$_{1}$ (left) and Q$_{4}$ (right) readout IQ histograms used in state discrimination. The relatively low signal to noise ratio in the setup limits our ability discriminate the qubit states with high fidelity.}
    \label{fig:readoutcorr}
\end{figure}

As discussed in Sec.~\ref{sec:2qgatecal} the device parameters related to readout quality and signal to noise ratio (SNR) are not optimal in the device used in this work \cite{BlaisCQED2004,Blais2020Review}. Example state discrimination is shown in Fig.~\ref{fig:readoutcorr}\textbf{b}. To mitigate measurement errors that result from this we perform joint readout correction on the raw measured populations. We first calibrate the readout confusion matrix, $C_{ij} = P(i|j)$, which is the matrix of probabilities of measuring the state $i$ given that the true state (prepared state) is $j$, where $i$ denotes the state $\ket{i_{1},\dots,i_{N}}$ for $i_{n} = 0,1$ on a subset of $N$ of the device qubits. This matrix is inverted and multiplied by the vector of measured state probabilities, $p_{j}$, to yield the readout error corrected state probabilities  $q_{i}=C_{ij}^{-1}p_{j}$, which are displayed throughout the results in the main text. In Fig.~\ref{fig:readoutcorr}\textbf{a} we plot the elements of the two qubit confusion matrix $C_{ii}$ for Q$_{1}$ and Q$_{4}$ and note that the probabilities of correctly reading out a given state, $P(i|i)$ are only 0.30 to 0.60. Further in Fig.~\ref{fig:readoutcorr}\textbf{a} we show the fluctuations of these probabilities across 16 $\cross$ 30 iterations of the confusion matrix measurements (interleaved in the QPT measurements in Sec.~\ref{sec:2qgatecal}) over a 12-hour time period. As discussed in the main text, these joint readout fluctuations are likely responsible for most of the remaining process infidelity of the calibrated fSim gate after accounting for the contribution from decoherence. 

\section{Device Parameters and Driven Decoherence\label{sec:devicepars}}
\begin{table*}
    \centering
    \begin{tabular}{|c|c|c|c|c|c|}
        \hline
        Qubit & 1 & 4 & 6 & 7 & 8 \\
        \hline \hline
        max. $\omega_{01}/2\pi$ (GHz) & 5.347 & 5.432 & 5.801 & 5.882 & 5.831 \\[2pt]
        min. $\omega_{01}/2\pi$ (GHz) & 4.681 & 4.698 & 5.072 & 5.112 & 5.088 \\[2pt]
        anharmonicity (MHz) & 217 & 227 & 264 & 263 & 262 \\[2pt]
        $T_{1}$ ($\mu$s) & 10.8 $\pm$ 1.1 & 13.3 $\pm$ 3.2 & 13.4 $\pm$ 3.6 & 12.0 $\pm$ 1.4 & 12.7 $\pm$ 3.6 \\[2pt]
        $T_{2r}$ ($\mu$s) & 5.3 $\pm$ 0.24 & 4.7 $\pm$ 0.20 & 3.9 $\pm$ 0.41 & 4.4 $\pm$ 2.8 & 4.3 $\pm$ 0.56 \\[2pt]
        $T_{2e}$ ($\mu$s) & 7.2 $\pm$ 0.34 & 11.6 $\pm$ 1.0 & 8.7 $\pm$ 1.2 & 6.3 $\pm$ 2.1 & 12.7 $\pm$ 3.6 \\[2pt]
        $T_{1,\text{driven}}$ ($\mu$s) & 10.2 $\pm$ 3.9 & 6.9 $\pm$ 0.80 & - & - & - \\[2pt]
        $T_{2r,\text{driven}}$ ($\mu$s) & 3.4 $\pm$ 0.23 & 3.2 $\pm$ 0.19 & - & - & - \\[2pt]
        $T_{2e,\text{driven}}$ ($\mu$s) & 3.1 $\pm$ 0.14 & 3.0 $\pm$ 0.21 & - & - & - \\[2pt]
        $g_{0i}/2\pi$ (MHz) & 18 & 17 & 15 & 15 & 15 \\
        \hline
    \end{tabular}
    \caption{Device parameters for the five qubits used in the measurements presented throughout the main text. Minimum and maximum qubit frequencies and qubit anharmonicities were determined from two-tone spectroscopy and qubit-Bus couplings $g_{0i}$ were determined from spectroscopy of qubit-Bus avoided crossings (see Sec.~\ref{sec:freqdomain}).}
    \label{tab:devicepars}
\end{table*}
Device parameters for the five qubits used for the measurements in the main text are shown in Table~\ref{tab:devicepars}. Coherence time measurements were each repeated 50 times over a 12-hour period to capture the effects of temporal fluctuations in the energy decay time $T_{1}$, Ramsey dephasing time $T_{2r}$, and Hahn-echo dephasing time $T_{2e}$. We observe that modulation of the Bus boundary flux bias introduces additional decoherence in the system. As a result we similarly measure $T_{1,\text{driven}}$, $T_{2r,\text{driven}}$, and $T_{2e,\text{driven}}$ for Q$_{1}$ and Q$_{4}$ in the presence of a Bus flux modulation drive at the same amplitude used for the $\text{fSim}$ gate between qubits Q$_{1}$ and Q$_{4}$ (see Sec.~\ref{sec:2qgatecal}), but detuned from the gate frequency by $+20~\text{MHz}$ to prevent coherent population exchange between the qubits during the coherence time measurements. These results are used for estimating the coherence limited process fidelity, $\mathcal{F}_{\text{coh}}$, in Sec.~\ref{sec:2qgatecal}. 

\section{Frequency Constraints and Allocation \label{sec:freqalloc}}
Extending the result in Sec.~\ref{sec:tswt} to the case of $N$ qubits with tunable coupling to a shared Bus resonator a re-configurable quantum processor (QPU) can be implemented. Re-configurable refers to the fact that this architecture allows one to implement any arbitrary connectivity graph between the $N$ qubits. The potential for all-to-all coupling among the qubits is enabled by utilizing one shared Bus resonator to couple to all the qubits. The tunability of the coupling and the spectrally selective nature of the parametric coupling processes help mitigate the unmanageable crosstalk that is typically present in high connectivity systems of qubits. Achieving full programmability places constraints on the allocation of frequencies in the system. The number of frequency constraints grows as $N$ increases. In practice, this requirement on the spectrum of qubit-qubit detunings imposes a trade-off between the number of qubits $N$ and the effective coupling rate $g_{\text{eff}}$ given a finite bandwidth for the qubit frequencies.  We need to avoid the scenario where the coupling modulation at $\Delta_{ij}$ generates unintentional population exchange in another qubit pair $kl$, one form of undesirable crosstalk which could be present in such a highly connected system. Intuitively we need to be able to sufficiently resolve all of the pairwise detunings between qubits so they can be addressed independently. The constraint can be written as
\begin{equation}
    g_{\text{eff}} \ll s_{\text{min}}= \min_{(i,j),(k,l)}|\Delta_{ij}-\Delta_{kl}|
    \label{eqn:allocation}
\end{equation}
where $(i,j)$ and $(k,l)$ enumerate the different pairs of qubits that one wishes to couple together and $g_{\text{eff}}$ is the maximum effective coupling strength allowed by a given frequency configuration. From the measurements in Fig.~\ref{fig:q14exchange} we found $g_{\text{eff}} \leq 4s_{\text{min}}$ is generally sufficient to keep the errors due to the presence of other qubits small. We now provide proof that a frequency allocation can be found which allows for $g_{\text{eff}}/2\pi \sim 5-9 \text{MHz}$ in a fully connected 8 qubit QPU. Larger $g_{\text{eff}}$ is of course possible if the degree of desired connectivity is reduced or larger quantum crosstalk errors can be tolerated. 

\begin{figure}
    \centering
    \includegraphics{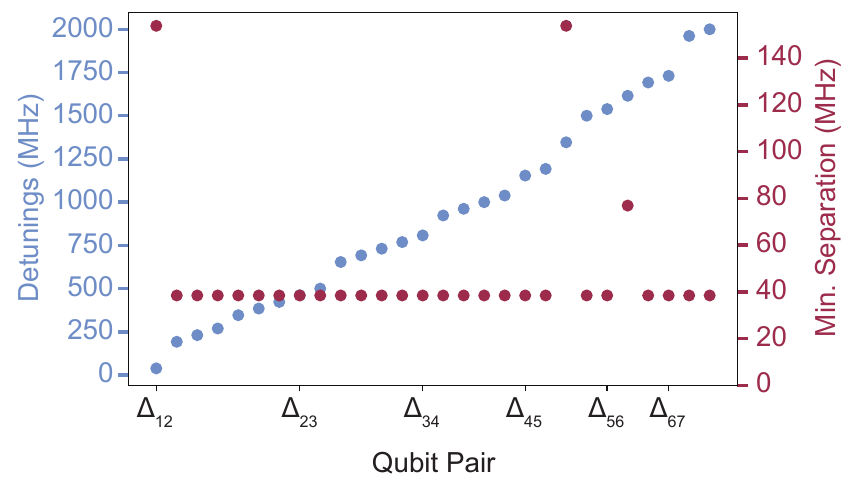}
    \caption{Results of 8 qubit frequency allocation for the proposed architecture. The pairwise detunings (blue, left axis), $\Delta_{ij}$, are plotted for each pair of qubits in the system after optimizing their frequencies based on Eqn.~\ref{eqn:allocation}. The detunings are sorted in increasing order for convenience. Additionally, the minimum detuning separation (red, right axis), $\min_{k,l}|\Delta_{ij}-\Delta_{kl}|$, is shown for each pair $(i,j)$. We find $s_{\min}/2\pi=35~\text{MHz}$ which allows for effective coupling rates of $g_{\text{eff}}/2\pi \approx 5-9~\text{MHz}$ with qubits spread across a $2\pi\times2~\text{GHz}$ band centered on the bus frequency.}
    \label{fig:freqalloc}
\end{figure}

We assume that we work with the fourth order mode of a linear resonator with $\omega_{1}/2\pi=1.5~\text{GHz}$ so the mode of interest has $\omega_{\text{r}}/2\pi=6~\text{GHz}$. Future devices using the achritecture outlined in this work can operate in this way while still coupling 8 qubits symmetrically to the 4th order Bus mode. Meanwhile, the larger FSR allows for larger qubit frequency bandwidth and as a result, larger $s_{\text{min}}$. In order to minimize the effects of the neighboring third and fifth order modes at $\omega_{3}/2\pi=4.5~\text{GHz}$ and $\omega_{5}/2\pi=7.5~\text{GHz}$ we limit the qubit frequencies to be placed within a $w/2\pi = 2~\text{GHz}$ bandwidth around $\omega_{\text{r}}$. This maintains a detuning of at least $500~\text{MHz}$ between each qubit and the neighboring Bus modes. Finally, to remain in the dispersive regime we require $\Delta_{i} \gg g_{0,i}$ for each qubit $i$ where $\Delta_{i}=\omega_{i}-\omega_{\text{r}}$ is the detuning of qubit $i$ from the Bus resonance and $g_{0i}$ is the maximum coupling of qubit $i$ to the Bus. Thus, we require $\min_{i}|\Delta_{i}| \geq 2\pi\times150~\text{MHz}$. It is straightforward to find a configuration that maximizes $s_{\text{min}}$ numerically given the constraints outlined above. For one such frequency configuration, we plot the Q$_{i}$, Q$_{j}$ detuning, $\Delta_{ij}$ for all qubit pairs, along with the minimum separation, $\min_{(k,l)}|\Delta_{ij}-\Delta_{kl}|$, of detuning $\Delta_{ij}$ from all other detunings in the system in Fig.~\ref{fig:freqalloc}. The minimum separation between any pair of detunings in this configuration is found to be $s_{\text{min}}/2\pi=35.3~\text{MHz}$ allowing for an effective coupling up to $g_{\text{eff}}/2\pi \leq \times 8.82~\text{MHz}$ or for stronger mitigation of errors by choosing a smaller $g_{\text{eff}}$.

In practice, the frequency precision required for this allocation to be successful is not possible in fixed frequency transmons given state of the art Josephson junction uniformity \cite{Kreikebaum_2020}. In the future, more advanced frequency allocation methods \cite{morvan2021optimizing} in combination with post fabrication laser annealing \cite{ZhangLaser2022,Hertzberg2021,KimLaser2022} may allow a fixed frequency implementation. In the near term, as in this work, tunable transmons would be required. This would require implementation of engineering advances to improve the qubit coherence times away from the flux insensitive points so one can take advantage of the full tuning range of the qubits \cite{Hutchings2017}.

\section{Extended Frequency Domain Data \label{sec:extdataFD}}
\begin{figure}
    \centering
    \includegraphics{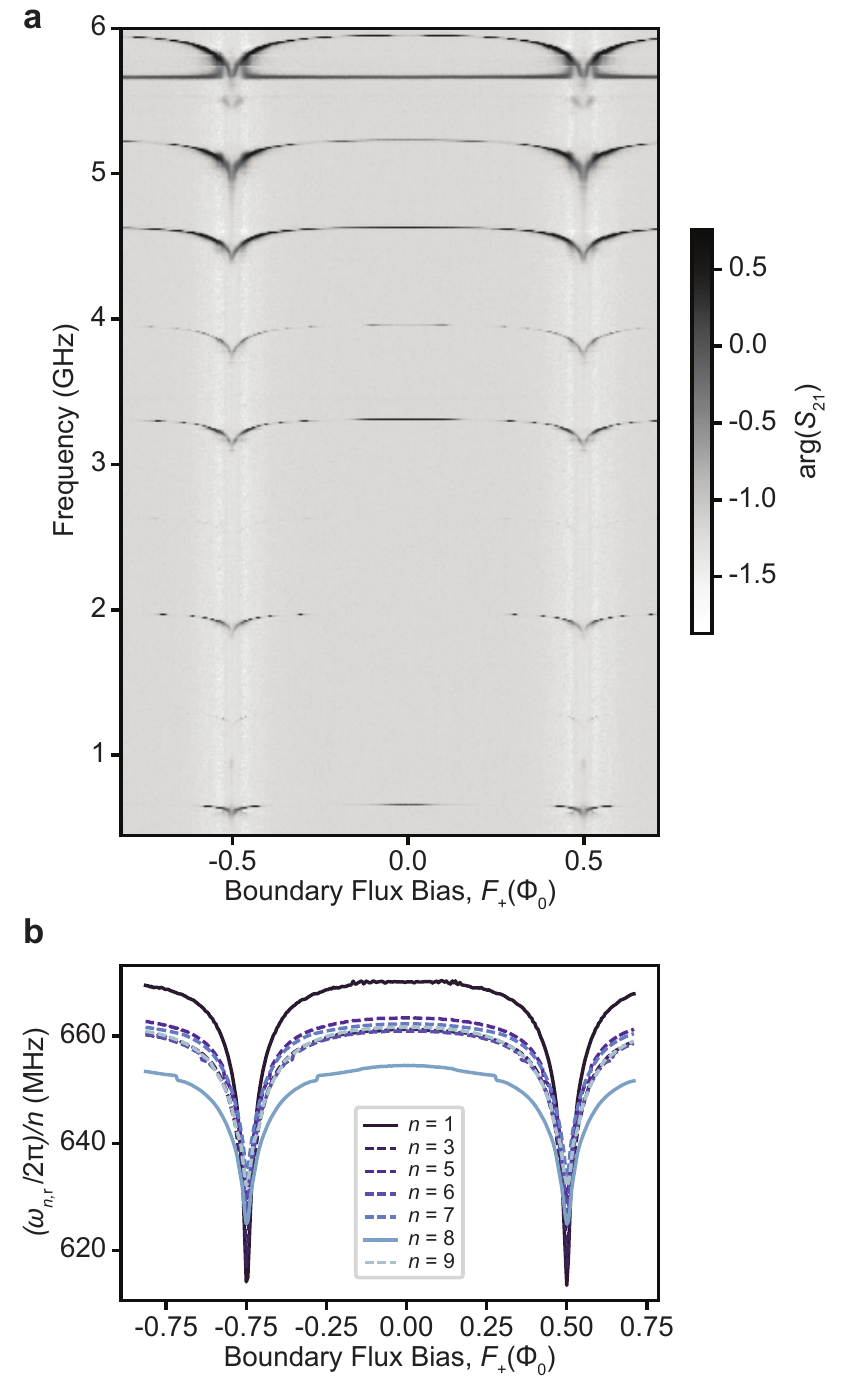}
    \caption{\textbf{a}, Bus spectroscopy of first 9 resonant modes as a function of boundary flux bias. The data in Fig.~\ref{fig:spidernetcircuit}\textbf{b} is a subset of this data from 3.1-4.65 GHz. The color scale of the full data reproduced here is modified to visually emphasize the resonant features. The flat feature near the 9th order mode corresponds to the frequency of the qubit (Q$_{5}$) used as a spectrometer for these measurements. \textbf{b}, Normalized mode frequencies as a function of boundary flux bias, fitted from the full spectroscopy data in \textbf{a}. Solid curves show modes which are excluded from the fit outlined in the main text.}
    \label{fig:extendedbusspectrum}
\end{figure}
We will now present extended data sets from the FD measurements in the main text and described in Sec.~\ref{sec:freqdomain}. First in Fig.~\ref{fig:extendedbusspectrum}\textbf{a} we show Bus spectroscopy data over the full frequency range probed. Here we see clear signatures (dark features) of modes $n=1,3,5,6,7,8,9$ at frequencies $\omega_{n,\text{r}}$, highlighting the multi-mode structure of the Bus. Additionally, there are faint features, most visible near $F_{+}=\pi/2$, corresponding to the $n=2,4$ modes which only couple weakly to the qubit used as a spectrometer for these measurements. Since these modes were not detected at all $F_{+}$ in this measurement, we exclude them from the multi-mode fit used to determine the circuit parameters in Table~\ref{tab:DCfitpars}. Below the spectroscopy data, in Fig.~\ref{fig:extendedbusspectrum}\textbf{b}, we plot the mode frequencies normalized by the mode index, $\omega_{n,\text{r}}$. This highlights that modes $n=1,8$ are outliers in terms of their frequencies. For mode $n=8$, which is closest in frequency to the 8 qubits and has anti-nodes at their positions when $F=0$, we suspect that Lamb shifts (see Eqn.~\ref{eqn:fulltransformH}) due to the Bus coupling to the qubits are responsible for the overall shift down. For mode $n=1$, the shift may be attributable to the comparatively higher energy participation of this mode in the boundary SQUID capacitances. As a result we also exclude modes $n=1,8$ from the fit.

\begin{figure}
    \centering
    \includegraphics{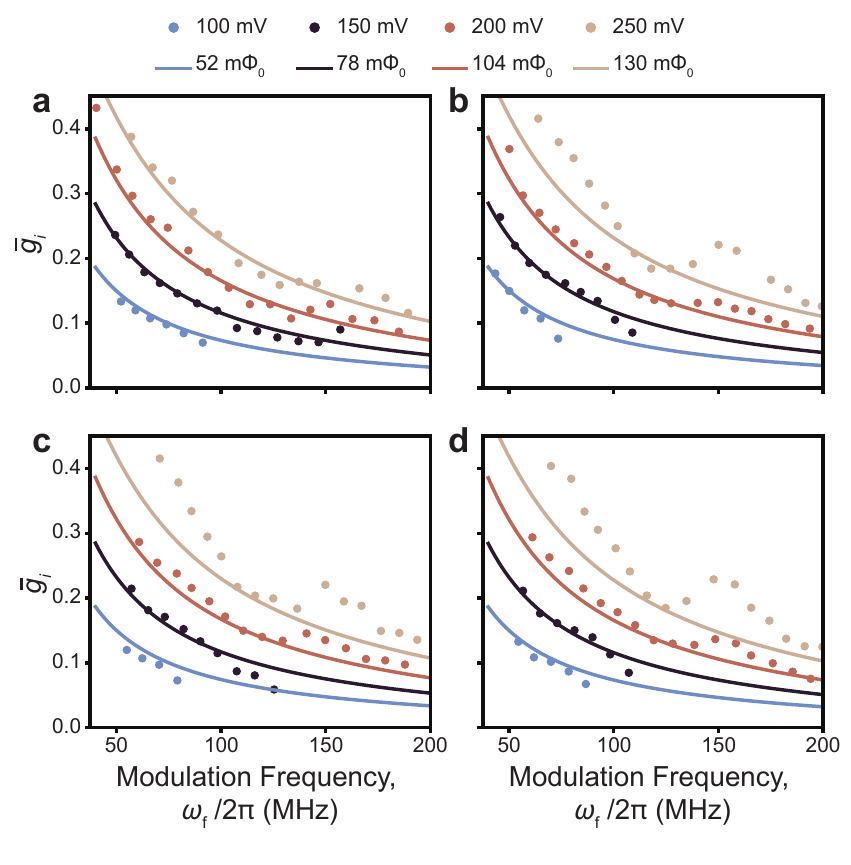}
    \caption{Full parametric coupling data for \textbf{a} Q$_{1}$, \textbf{b} Q$_{6}$, \textbf{c} Q$_{7}$, and \textbf{d} Q$_{8}$. In the main text, Fig.~\ref{fig:parametricqubitbus}\textbf{a}, we show the corresponding data for Q$_{4}$ as a representative example.}
    \label{fig:extendedgpar}
\end{figure}

Next in Fig.~\ref{fig:extendedgpar}\textbf{a}-\textbf{d} we show parametric qubit-Bus coupling rates measured in FD for Q$_{1}$, Q$_{6}$, Q$_{7}$, and Q$_{8}$ which can be compared to the results in Fig.~\ref{fig:parametricqubitbus} for Q$_{4}$. The theory curves plotted alongside the data are generated from the same set of parameters for all qubits and no further fitting is done for each particular qubit. Again, consistency between measurements on different qubits demonstrates that they couple symmetrically to the Bus as designed and strong agreement with the theory is obtained without fine-tuning any qubit specific parameters. 

\clearpage
\newpage

\section{Wiring \label{sec:wiring}}
\begin{figure*}[h]
    \centering
    \includegraphics[scale=0.92]{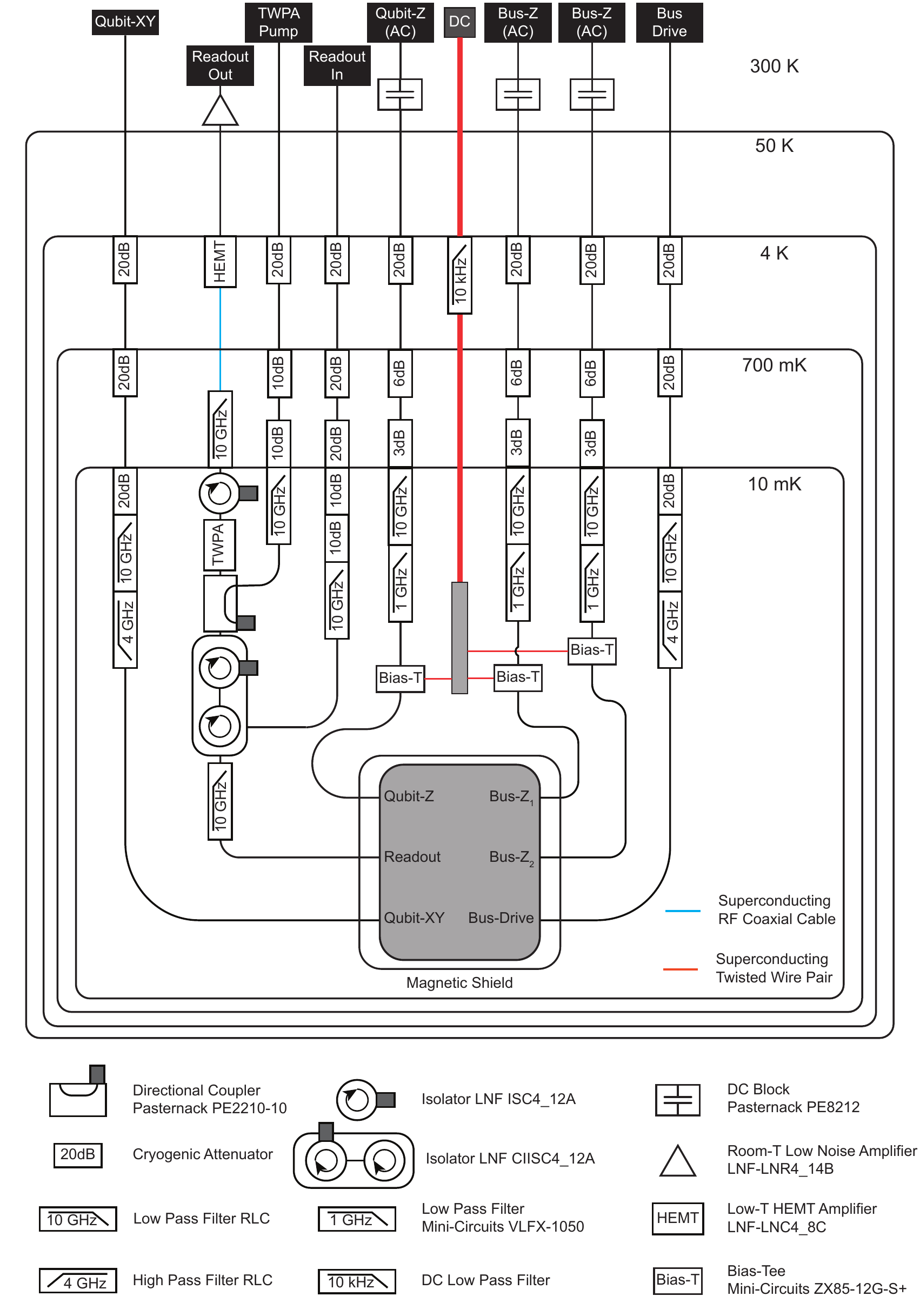}
    \caption{Experimental setup.}
    \label{fig:wiring}
\end{figure*}

The wiring scheme and experimental configuration used to measure the performance of the device for implementing programmable coupling between qubits through parametric modulation of the Bus resonator is shown in Fig.~\ref{fig:wiring}. Due to the space limitation, we only show one set of wiring for interfacing with one qubit in the figure. The number of qubit control lines (Qubit-Z and Qubit-XY) should be proportional to the number of qubits. In this work, we have 8 such sets of identical control lines.


\end{document}